\documentclass{article}
\usepackage{dsfont,amscd,amsmath,amssymb}
\def\DH{\rm I\kern-1.5pt\rm H\kern-1.5pt\rm I}

\newcommand{\bZ}{{\overline Z}}

\newcommand{\bQ}{{\overline Q}}
\newcommand{\bS}{{\overline S}}
\newcommand{\bK}{{\overline K}}
\newcommand{\bD}{{\overline D}}

\def\DR{\rm I\kern-1.45pt\rm R}
\def\DC{\kern2pt {\hbox{\sqi I}}\kern-4.2pt\rm C}

\newcommand{\mpsi}{\boldsymbol{\psi}}
\newcommand{\mbpsi}{{\bar{\boldsymbol\psi}}}
\newcommand{\bLambda}{{\overline \Lambda}}
\newcommand{\blambda}{{\bar\lambda}}
\newcommand{\mlambda}{\boldsymbol{\lambda}}
\newcommand{\mblambda}{\bar{\boldsymbol{\lambda}}}
\newcommand{\mLambda}{\boldsymbol{\Lambda}}
\newcommand{\mbLambda}{\overline{\boldsymbol{\Lambda}}}
\newcommand{\ml}{\boldsymbol{l}}
\newcommand{\mbl}{\bar{\boldsymbol{l}}}
\newcommand{\mT}{\boldsymbol{T}}
\newcommand{\mbT}{\overline{\boldsymbol{T}}}
\newcommand{\mY}{\boldsymbol{Y}}

\newcommand{\bV}{{\overline V}}
\newcommand{\bnabla}{{\overline{\nabla}}}

\newcommand{\cF}{{\cal F}}

\newcommand{\cD}{{\cal D}}
\newcommand{\cG}{{\cal G}}

\newcommand{\bpsi}{{\bar\psi}}

\newcommand{\nn}{\nonumber}
\newcommand{\ba}{\begin{array}}
\newcommand{\ea}{\end{array}}
\newcommand{\be}{\begin{equation}}
\newcommand{\ee}{\end{equation}}
\newcommand{\bea}{\begin{eqnarray}}
\newcommand{\eea}{\end{eqnarray}}
\newcommand{\bi}{\begin{itemize}}
\newcommand{\ei}{\end{itemize}}

\newcommand{\eps}{\varepsilon}

\newcommand{\p}[1]{(\ref{#1})}
\def\theequation{\arabic{section}.\arabic{equation}}
\begin{document}\thispagestyle{empty}
\thispagestyle{empty}
\begin{flushright}
\end{flushright}
\vspace{4cm}
\begin{center}
{\Large\bf Coset approach to the partial breaking of global supersymmetry}
\end{center}
\vspace{1cm}

\begin{center}
{\large\bf S.~Bellucci${}^a$, S.~Krivonos${}^{b}$,
A.~Sutulin${}^{a,b}$ }
\end{center}

\begin{center}
${}^a$ {\it
INFN-Laboratori Nazionali di Frascati,
Via E. Fermi 40, 00044 Frascati, Italy} \vspace{0.2cm}

${}^b$ {\it
Bogoliubov  Laboratory of Theoretical Physics, JINR,
141980 Dubna, Russia} \vspace{0.2cm}

\end{center}
\vspace{2cm}

\begin{abstract}\noindent
We propose a method  to construct the on-shell component actions for the theories with $1/2$ partial
breaking of global supersymmetry within the nonlinear realization (coset) approach. In contrast with
the standard superfield approach in which unbroken supersymmetry plays the leading role,  we have shifted
the attention to the spontaneously broken supersymmetry. It turns out that in the theories in which one half
of supersymmetries is spontaneously broken, all physical fermions are just the fermions of the nonlinear
realization. Moreover, the transformation properties of these fermions with respect to the broken supersymmetry
are the same as in the famous Volkov-Akulov model. Just this fact completely fixed all possible appearances of
the fermions in the component action: they can enter the action  through the determinant of the vielbein
(to compensate the transformation of the volume form) and the covariant derivatives, only.
It is very important that in our parametrization of  the coset the rest of physical components,
i.e. all  bosonic components, transform as  ``matter fields'' with respect to the broken supersymmetry.
Clearly, in  such a situation the component action acquires the form of the Volkov-Akulov action for these
``matter fields''. The complete form of the action can be further fixed by two additional requirements:
a) to reproduce the bosonic limit, which is explicitly known in many interesting cases, and b) to have a proper
linearized form, which has to be invariant with respect to the linearized unbroken supersymmetry.
In some cases the additional
Wess-Zumino terms (which disappear  in the bosonic limit) have to be added to the action.
We supply the general consideration by the detailed examples of the actions for the superparticle in $D=3$
realizing $N=4\cdot 2^{k} \rightarrow N=2\cdot 2^k$ pattern of supersymmetry breaking, the superparticle
in $D=5$ with the $N=16$ supersymmetry broken down to $N=8$ one, the on-shell component action
for $N=1, D=5$ supermembrane and its dual cousins and the component action of $N=1$ supermembrane in $D=4$.
In these cases we provide  the  exact proof of the invariance of the constructed component actions with respect
to both broken and unbroken supersymmetries.
\end{abstract}

\newpage
\setcounter{page}{1}
\setcounter{equation}{0}
\section{Introduction}
The characteristic feature of the theories with a partial breaking
of global supersymmetries is the appearance of the Goldstone
fermionic fields, associated with the broken supertranslations, as
the components of Goldstone supermultiplets of unbroken
supersymmetry. The natural description of such theories is
achieved within the coset approach \cite{coset1, coset2,VA}. The usefulness
of the coset approach in the applications to the theories with
partial breaking of the supersymmetry have been demonstrated by
many authors [4-20]. The presence of the unbroken supersymmetry
makes quite reasonable the idea to choose the corresponding
superfields as the basic ones, and many interesting superspace
actions describing different patterns of supersymmetry breaking
have been constructed in such a way \cite{BG1, RT, R2,IK}. However,
the standard methods of coset approach fail to construct the
superfield action, because the superspace Lagrangian is weakly
invariant with respect to supersymmetry - it is shifted by the
full space-time or spinor derivatives under broken/unbroken
supersymmetry transformations. Another rather technical
difficulty is the explicit construction of the component action from the superspace
one, which is written in terms of the superfields subjected to
highly nonlinear constraints. Finally, in some cases the
covariantization of the irreducibility constraints with respect to
the broken supersymmetry is not evident, if at all possible. For
example, it has been demonstrated in \cite{BG1} that such
constraints for the vector supermultiplet can be covariantized
only together with the equations of motion.

It turned out that one can gain more information about component
off-shell actions if an attention is shifted to the broken
supersymmetry. It was demonstrated in~\cite{BKS1},~\cite{BKKS1},~\cite{BKKY} that with a
suitable choice of the parametrization of the coset, the
$\theta$-coordinates of unbroken supersymmetry and the physical
bosonic components do not transform under broken supersymmetry.
Moreover, the physical fermions transform as the Goldstino of the
Volkov-Akulov model \cite{VA} with respect to broken
supersymmetry. Therefore, the physical fermions can enter the
component on-shell action only i) through the determinant of the vielbein
(to compensate the variation of the volume $d^d x$),
ii) through the covariant space-time derivatives, or iii)
through the Wess-Zumino term, if it exists. The first two
ingredients can be easily constructed within the coset method,
while the Wess-Zumino term can be also constructed from Cartan forms
following the recipe of ref.~\cite{HM}. As a result, we will have
the Ansatz for the action with several constant parameters, which
have to be fixed by the invariance with respect to unbroken
supersymmetry. The pleasant feature of such an approach is that
the fermions are ``hidden'' inside the covariant derivatives and
determinant of the vielbein, making the whole action short, with
the explicit geometric meaning of each term. In the present paper
we review this procedure in applications to the actions of the superparticle in $D=3$ realizing
$N=4\cdot 2^{k} \rightarrow N=2\cdot 2^k$ pattern of supersymmetry breaking, the action of superparticle
in $D=5$ with the $N=16$ supersymmetry broken down to $N=8$ one, the on-shell component actions
of $N=1, D=5$ supermembrane and its dual cousins and the component action of $N=1$ supermembrane in $D=4$.
All these explicit actions confirm our conjecture about the structure of the component action.
Finally, we briefly discuss some related questions and further possible applications of our method.

\setcounter{equation}{0}
\section{Basics of the method}
In this section we present main features of the coset approach, applying to supersymmetric models in which one half
of the global supersymmetries are spontaneously broken. Before going to supersymmetric systems, we will consider
how this method works in the purely bosonic case.

Let us split the generators of the target space of the $D$-dimensional Poincar\'{e} group, which is supposed
to be spontaneously broken on the world volume down to the $d$-dimensional Poincar\'{e} subgroup, into the generators
of unbroken $\{P , M, N\}$ and spontaneously broken
$\{Z, K\}$ symmetries. The generators $P$ and $Z$ form $D$-dimensional translations, $M$ generators span the
$so(1,d-1)$ - Lorentz algebra on the world volume, the generators $N$ rotate broken translations $Z$ among
themselves and thus they span $so(D-d)$ algebra,  while generators $K$ belong to the coset $so(1,D-1)/so(1,d-1)\times so(D-d)$.
All transformations of the $D$-dimensional Poincar\'{e} group can be realized by the left action of different group elements on the coset space
\footnote{For the sake of brevity we suppress here all space-time indices.}
\be\label{I1}
g= e^{x P} e^{q(x) Z} e^{\Lambda(x)K}.
\ee
The spontaneous breaking of $Z$ and $K$ symmetries is reflected in the character of corresponding coset coordinates
which are Goldstone fields $q(x)$ and $\Lambda(x)$ in the present case. The transformation properties of coordinates
$x$ and fields $\{q(x), \Lambda(x)\}$ may be easily found in this approach, while all needed information about the geometry of the coset space \p{I1}
is contained in the Cartan forms
\be\label{I2}
g^{-1}d g = \Omega_P P + \Omega_M M+ \Omega_Z Z+\Omega_K K+ \Omega_N N.
\ee
All Cartan forms except for $\Omega_M$ and $\Omega_N$ are transformed homogeneously under all symmetries.
Due to the general theorem \cite{ih} not all of the above Goldstone fields have to be treated as independent. In the present
case the fields $\Lambda(x)$ can be covariantly expressed through $x$-derivatives of $q(x)$ by imposing the constraint
\be\label{I3}
\Omega_Z=0.
\ee
Equations encoded in the conditions \p{I3}, do not contain dynamic restrictions and are purely kinematic.
Thus, we are dealing with the fields $q(x)$ only. It is very important that the form $\Omega_P$ defines the vielbein
$E$ ($d$-bein  in the present case), connecting the covariant world volume coordinate differentials $\Omega_P$ and
the world volume coordinate differential $dx$ as
\be\label{I4e}
\Omega_P = E \cdot dx .
\ee
Combining all these ingredients, one may immediately write the  action
\be\label{I4}
S = -\int d^dx + \int d^dx\; \det E,
\ee
which is invariant under all  symmetries. In \p{I4} we have added the  trivial first term to fulfill  the condition $S_{q=0}=0$.
The action \p{I4} is just the static gauge form of the action of $p=(d-1)$-branes.

The supersymmetric generalization of the coset approach involves into the game new spinor generators $Q$ and $S$ which extend the
$D$-dimensional Poincar\'{e} group to the supersymmetric one
\be\label{II1}
\left\{ Q,Q\right\} \sim P, \quad \left\{ S,S\right\} \sim P, \quad \left\{ Q,S\right\} \sim Z.
\ee
The most interesting cases are those  when the $Q$ supersymmetry is kept unbroken, while the $S$ supersymmetry is supposed
to be spontaneously broken\footnote{If all supersymmetries are considered as spontaneously broken, the corresponding
action can be constructed similarly to the bosonic case, resulting in the some synthesis of Volkov-Akulov \cite{VA}
and Nambu -Goto actions. An enlightening example of such a construction can be found in \cite{CNV}.}. When $\#Q=\#S$
we are facing the so-called $1/2$ Partial Breaking of Global Supersymmetry cases  (PBGS),  which most of all interesting
supersymmetric domain walls belong to. Only such cases of supersymmetry breaking will be considered in this paper.

Now, all symmetries can be realized by group elements acting on the coset element
\be\label{I6}
g= e^{x P} e^{\theta Q} e^{\boldsymbol{q}(x,\theta) Z} e^{\boldsymbol{\psi}(x,\theta)S} e^{\boldsymbol{\Lambda}(x,\theta)K}.
\ee
The main novel feature of the supersymmetric coset \p{I6} is the appearance of the Goldstone superfields
$\{\boldsymbol{q}(x,\theta)$, $\boldsymbol{\psi}(x,\theta)$, $\boldsymbol{\Lambda}(x,\theta)\}$ which depend on the coordinates
of the world volume superspace $\{x, \theta\}$.
The rest of the coset approach machinery works in the same manner: one may construct the Cartan forms \p{I2} for the coset \p{I6}
(which will contain the new forms $\Omega_Q$ and $\Omega_S$), one may find the supersymmetric $d$-bein and corresponding
bosonic $\nabla_P$ and spinor $\nabla_Q$ covariant derivatives, etc. One may even  write the proper generalizations
of the covariant constraints \p{I3} as
\be\label{I7}
\Omega_Z=0, \quad \Omega_S|=0,
\ee
where $|$ means the $d\theta$-projection of the form (see e.g.~\cite{BIK1} and references therein).
The $d\theta$-parts of these constraints are closely related with the "geometro-dynamical" constraint of the superembedding approach (see e.g.~\cite{Dima}).

Unfortunately, this similarity between purely bosonic and supersymmetric cases is not complete due to the existence of
the following important new features of theories with partial breaking of global supersymmetry:
\begin{itemize}
\item In contrast with the bosonic case, not all of the physical fields appear among the parameters of the coset.
A famous example comes from the supersymmetric space-filling ${\rm D3}$-brane (aka $N=1$ Born-Infeld theory) where the coset element \p{I6}
contains only $P$, $Q$ and $S$ generators \cite{BG2, RT}, while the field strength $F$ is ``hidden'' inside
    the superfield $\boldsymbol{\psi}: F\sim \nabla_Q \boldsymbol{\psi}|$.
    Nevertheless, {\it it is true} that the {\it all physical bosonic components}
    can be found in the  quantity $\nabla_Q \boldsymbol{\psi}|$.
\item The supersymmetric generalization \p{I7} of the bosonic kinematic constraints \p{I3} in most cases contains
not only kinematic conditions, but also dynamic superfield equations of motion. A prominent example again may be found
in \cite{BG2}. Moreover, in many  cases it is unknown how to split these constraints into kinematical and
dynamical ones.
\item But the most unpleasant feature of the supersymmetric cases is that the standard methods of nonlinear realizations
fail to construct the superfield action! The main reason for this is simple: all that we have at hands are the covariant
Cartan forms,  which we can construct the  superfield invariants from, while the superspace Lagrangian is not invariant.
Instead it is shifted by the full spinor derivatives under unbroken and/or broken supersymmetries.
\end{itemize}

Nevertheless, we are going to apply a coset approach to the supersymmetric cases and to demonstrate
how on-shell component actions can be constructed within it.
The main idea is to start  with the Ansatz for the action  manifestly invariant with respect to
{\it spontaneously broken supersymmetry}. Funny enough, it is rather easy to do, due to the following properties:
\begin{itemize}
\item In our parametrization of the coset element \p{I6} the superspace coordinates $\theta$ do not transform
under broken supersymmetry. Thus, all components of superfields transform {\it independently},
\item The covariant derivatives $\nabla_P$ and $\nabla_Q$ are invariant under broken supersymmetry.
Therefore, the bosonic physical components which are contained in $\nabla_Q \boldsymbol{\psi}(x,\theta)| $ can be treated as  ``matter fields''
    (together with the field $\boldsymbol{q}(x,\theta)|$ itself)
    with respect to broken supersymmetry,
\item All physical fermionic components are just $\theta=0$ projections of the superfield $\boldsymbol{\psi}(x, \theta)$ and
these components transform as the fermions of the Volkov-Akulov model \cite{VA} with respect to broken supersymmetry.
\end{itemize}
The immediate consequence of these facts is the conclusion that the physical fermionic components can enter the component on-shell
action either through the determinant of the $d$-bein constructed with the help of the Cartan form $\Omega_P$ in the limit $\theta=0$,
namely, ${\cal E} = E|$,
through the space-time derivatives of the ``matter fields'' $\nabla_P \boldsymbol{q}|$, or through the Wess-Zumino
terms if they exist. Thus, the most general Ansatz for
the on-shell component action, which is invariant with respect to spontaneously broken supersymmetry, has the form
\be\label{I8}
S=\int d^d x - \int d^dx \det {\cal E} {\cal F}( \nabla_Q \boldsymbol{\psi}|, \nabla_P \boldsymbol{q}|)+S_{WZ}.
\ee
Note, that the arguments of the function ${\cal F}$ are the bosonic physical components $\nabla_Q \boldsymbol{\psi}|$ and
the covariant space-time derivatives of $\boldsymbol{q}$ (which, by the way, are also contained in $\nabla_Q \boldsymbol{\psi}|$).
In certain cases, for fixing an explicit form of the function ${\cal F}$ it is sufficient that the following two conditions be satisfied
\begin{enumerate}
\item The action \p{I8} should have a proper bosonic limit, which is known in almost all interesting cases.
One should note, that this limit for the action \p{I8} is trivial
$$ S_{bos}=\int d^d x\Big ( 1 -{\cal F}( \nabla_Q \boldsymbol{\psi}|, \partial_P q)\Big ).$$
\item The action \p{I8} in the linear limit should possess a linear version of unbroken supersymmetry, i.e. it should be just
the sum of the kinetic terms for all bosonic and fermionic components with the relative coefficients fixed by
unbroken supersymmetry.
\end{enumerate}
One should note that the Wess-Zimino action, which is invariant under broken supersymmetry,  can be also constructed from
the Cartan forms following the recipe of ref.~\cite{HM}. Thus, the role of unbroken supersymmetry is to fix
the coefficients in the action \p{I8} to achieve its invariance with respect to unbroken sypersymmetry.

In the next two sections of the present paper we will show how the coset approach works in the cases of the superparticle
in $D=3$ and $D=5$  with the chiral and quartet Goldstone supermultiplets, respectively.
Then in section 5 we will extend our analysis to the cases of $N=1$ supermembrane in $D=4$ as well as
of the dual system - $N=1$ supersymmetric space filling $\rm{D2}$-brane.
In section 6 we will show that in order to construct $N=2$ supersymmetric action for the supermembrane action in $D=4$,
one needs to add the corresponding Wess-Zumino term.
In Appendices we collect the technical details, notation and explicit proof of invariance of the supermembrane action
with respect to both, broken and unbroken supersymmetries. We conclude with some comments and perspectives.

\setcounter{equation}{0}
\section{Superparticle in D=3}
The main goal of this section is to provide the detailed structure of the component on-shell actions for the
one-dimensional system realizing a one half breaking of the global supersymmetry.
As an example, we consider a system
with $N=16 \rightarrow N=8$ pattern of supersymmetry breaking
based on the superalgebra with two "semi-central charges" $(Z, \overline Z)$.
We show that the resulting component action describes a superparticle in $D=3$.

\subsection{Superparticle in D=3: kinematics}
It is a well known fact that the action for the given pattern of the supersymmetry breaking is completely defined by
the choice of the corresponding Goldstone supermultiplet~\cite{BG1,BG2,BG3,RT,R2,IK,BIK1,BIK2,BIK3}. The bosonic scalars
of the supermultiplet are associated with the "semi-central charges" in the supersymmetry algebra \p{II1}.
To describe a system with one complex boson (or two real bosons) one has to choose $N=16, d=1$ Poincar\'{e} superalgebra with two "semi-central charges" $(Z, \overline Z)$
\be\label{N16algebra}
\left\{ Q^{i a}, \overline Q_{j b}  \right\}=2\delta^a_b \delta^i_j P\,, \quad
\left\{ S^{i a}, \overline S_{j b}  \right\}=2\delta^a_b \delta^i_j P\,, \quad
\left\{ Q^{ia}, S^{j b}  \right\} =2 i \eps^{a b} \eps^{ij} Z\,, \quad
\left\{ \overline Q_{i a}, \overline S_{j b}  \right\} = - 2 i \eps_{ab} \eps_{ij} \overline Z\,.
\ee
Here $i,a =1,2$ refer to the indices of the fundamental representations of two commuting $SU(2)$ groups.
In \p{N16algebra} $P$ is the generator of one-dimensional translation, while $Q^{ia}, \bQ_{ia}$ and $S^{ia}, \bS_{ia}$ are the generators
of unbroken and spontaneously broken $N=8$ supersymmetries,
respectively.
As we already explain in the Introduction, in the coset approach the statement that $S$ supersymmetry and $(Z,\bZ)$ translations are spontaneously
broken is reflected in the structure of the element of the coset space
\be\label{cosetN16}
g=e^{i tP}\,e^{\theta_{i a} Q^{i a} + \bar \theta^{i a} \overline Q_{i a}}\, e^{i (\boldsymbol{q} Z+\bar{\boldsymbol{q}} \overline Z)}\,
e^{\mpsi_{i a} S^{i a} + \bar{\mpsi}^{i a} \overline S_{i a}}.
\ee
Once we state that the coordinates $\mpsi$ and $\boldsymbol{q}$ are the superfields depending on the $N=8, d=1$ superspace
coordinates $(t, \theta, \bar\theta)$, then we are dealing with the spontaneously breaking of the corresponding
symmetries. Thus, in our case we will treat $\mpsi(t,\theta,\bar\theta)$, $\boldsymbol{q}(t,\theta,\bar\theta)$ as $N=8, d=1$ Goldstone
superfields accompanying $N=16 \rightarrow N=8$ breaking of supersymmetry in one dimension.

The transformation properties of coordinates and superfields under both unbroken and broken supersymmetries
are induced by the left multiplications of the group element $g_0$ on the coset \p{cosetN16}
$$g_0\, g = g'\, .$$
Thus, for the unbroken supersymmetry with $g_0= e^{\eps_{i a} Q^{i a} + \bar \eps^{i a} \overline Q_{i a}}$
one gets
\be\label{N16tran-U}
\delta_Q t = i \left(\eps_{i a} \bar \theta^{i a} + \bar \eps^{i a} \theta_{i a}  \right), \quad
\delta_Q\theta_{ia} = \eps_{i a}, \quad \delta_Q \bar \theta^{i a} = \bar \eps^{i a},
\ee
while for the broken supersymmetry with $g_0=e^{\eta_{i a} S^{i a} + \bar{\eta}^{i a} \overline S_{i a}}$
the transformations read
\be\label{N16tran-B}
\delta_S t = i \big (\eta_{i a} \bar \mpsi^{i a} + \bar \eta^{i a} \mpsi_{i a}  \big ), \quad
\delta_S\mpsi_{i a} = \eta_{i a}, \quad \delta_S\bar \mpsi^{i a} = \bar \eta^{i a}, \quad
\delta_S \boldsymbol{q} = - 2 \eta_{i a} \theta^{i a}, \quad \delta_S \bar{\boldsymbol{q}} = 2 \bar \eta^{i a} \bar\theta_{i a}.
\ee

The local geometric properties of the system are specified by the left-invariant Cartan forms
\be\label{N16-Cartan}
g^{-1} dg = i \omega_P P + (\omega_{Q})_{i a}Q^{i a} + (\bar\omega_{Q})^{i a} \overline Q_{i a}
+ i \omega_Z Z + i \bar\omega_Z \overline Z
+ (\omega_{S})_{i a} S^{i a} +( \bar\omega_{S})^{i a} \overline S_{i a}
\ee
which can be explicitly written in the considered case as
\bea\label{N16-form}
&&\omega_P = dt -i \big (\bar \theta^{ia} d\theta_{ia}  + \theta_{ia} d \bar \theta^{ia}
+  \bar \mpsi^{ia} d\mpsi_{ia} + \mpsi_{ia} d \bar \mpsi^{ia} \big ), \quad (\omega_Q)_{ia} = d\theta_{ia}, \quad
(\bar\omega_{Q})^{ia} = d\bar \theta^{ia},\nn \\
&& (\omega_{S})_{ia} = d\mpsi_{ia},\quad (\bar\omega_{S})^{ia} = d\bar{\mpsi}^{ia},\quad
\omega_Z = d \boldsymbol{q} + 2\mpsi^{ia} d\theta_{ia},\quad \bar \omega_Z = d \bar{\boldsymbol{q}} - 2 \bar \mpsi_{ia} d \bar \theta^{ia}.
\eea
Using the covariant differentials $(\omega_P, d\theta_{ia}, d \bar \theta^{ia})$ \p{N16-form},
one may construct the covariant derivatives
\bea\label{d1N16covder1}
&&
\partial_t = E\, \nabla_t \,, \quad E = 1- i \left( \mpsi_{ia} \dot{\bar\mpsi}{}^{ia}
+ \bar \mpsi^{ia} \dot \mpsi_{ia}  \right), \quad
E^{-1} = 1+ i \left( \mpsi_{ia} \nabla_t \bar \mpsi^{ia} + \bar \mpsi^{ia} \nabla_t \mpsi_{ia}  \right), \nn\\
&&
\nabla^{ia} = D^{ia} - i \left( \mpsi_{kb} D^{ia} \bar \mpsi^{kb} + \bar \mpsi^{kb} D^{ia} \mpsi_{kb} \right)\nabla_t
= D^{ia} - i \left( \mpsi_{kb} \nabla^{ia} \bar \mpsi^{kb} + \bar \mpsi^{kb} \nabla^{ia} \mpsi_{kb} \right)\partial_t, \nn \\
&&
\overline{\nabla}_{ia} = \overline{D}_{ia} - i \left( \mpsi_{kb} \overline{D}_{ia} \bar \mpsi^{kb}
+ \bar \mpsi^{kb} \overline{D}_{ia} \mpsi_{kb} \right)\nabla_t =
\overline{D}_{ia} - i \left( \mpsi_{kb} \overline{\nabla}_{ia} \bar \mpsi^{kb}
+ \bar \mpsi^{kb} \overline{\nabla}_{ia} \mpsi_{kb} \right)\partial_t\,,
\eea
where
\be\label{N16-flatDer}
D^{ia} = \frac{\partial}{\partial \theta_{i a}} -i \bar \theta^{i a} \partial_t, \quad
\overline D_{ia} = \frac{\partial}{\partial \bar \theta^{i a}} -i \theta_{i a} \partial_t,  \quad
\left\{ D^{ia} , \overline D_{jb} \right\}
= -2 i \delta^a_b \delta^i_j \partial_t.
\ee
The covariant derivatives \p{d1N16covder1} satisfy the following (anti)commutation relations
\bea\label{d1N16covder2}
&&
\left\{ \nabla^{ia} , \nabla^{jb} \right\}
= -2 i \left( \nabla^{ia} \mpsi_{kc} \nabla^{jb} \bar \mpsi^{kc}
+ \nabla^{ia} \bar \mpsi^{kc} \nabla^{jb} \mpsi_{kc} \right)\nabla_t,\nn \\
&&
\left\{ \overline{\nabla}_{ia} , \overline{\nabla}_{jb} \right\}
= -2 i \left( \overline{\nabla}_{ia} \mpsi_{kc} \overline{\nabla}_{jb} \bar \mpsi^{kc}
+ \overline{\nabla}_{ia} \bar \mpsi^{kc} \overline{\nabla}_{jb} \mpsi_{kc} \right)\nabla_t,\nn \\
&&
\left[ \nabla_t , \nabla^{ia} \right]
= -2 i \left( \nabla_t \mpsi_{kc} \nabla^{ia} \bar \mpsi^{kc}
+ \nabla_t \bar \mpsi^{kc}  \nabla^{ia} \mpsi_{kc} \right)\nabla_t, \nn \\
&&
\left[ \nabla_t , \overline{\nabla}_{ia} \right]
= -2 i \left( \nabla_t \mpsi_{kc} \overline{\nabla}_{ia} \bar \mpsi^{kc}
+ \nabla_t \bar \mpsi^{kc}  \overline{\nabla}_{ia} \mpsi_{kc} \right)\nabla_t, \nn \\
&&
\left\{ \nabla^{ia} , \overline{\nabla}_{jb} \right\}
= -2 i \delta^a_b \delta^i_j \nabla_t -2 i \left( \nabla^{ia} \mpsi_{kc} \overline{\nabla}_{jb} \bar \mpsi^{kc}
+ \nabla^{ia} \bar \mpsi^{kc} \overline{\nabla}_{jb} \mpsi_{kc}  \right)\nabla_t.
\eea

To reduce the number of independent Goldstone superfields let us impose the conditions on the $d\theta$-projections
of the Cartan forms $(\omega_Z, \bar \omega_Z)$ \p{N16-form}
\be\label{N16-kin}
\left\{
\begin{array}{l}
\omega{}_Z|_{\theta} =0, \\
\overline{\omega}_Z|_{\theta} = 0,
\end{array}\right.
\quad \Rightarrow \quad \left\{
\begin{array}{l}
\overline{\nabla}_{i a} \boldsymbol{q} =0, \quad \nabla^{i a} \boldsymbol{q} - 2 \mpsi^{i a} =0,\\
\nabla^{i a} \bar{\boldsymbol{q}} =0, \quad \overline{\nabla}_{i a} \bar{\boldsymbol{q}} + 2 \bar \mpsi_{i a} =0.
\end{array}\right.
\ee One part of these kinematical constraints can be recognized as
the covariant chirality conditions on the superfields
$\boldsymbol{q}$ and $\bar{\boldsymbol{q}}$, while the remaining
two equations express the fermionic Goldstone superfields
$\mpsi^{i a}$ and $\bar \mpsi_{i a}$ as the spinor derivatives of
the bosonic superfields $\boldsymbol{q}$ and
$\bar{\boldsymbol{q}}$, thereby realizing the Inverse Higgs
phenomenon~\cite{ih}. 

\subsection{Superparticle in D=3: dynamics}
It is well known that the standard chirality conditions are not enough to select an irreducible $N=8, d=1$ supermultiplet: one has impose
additional, second order in the spinor derivatives constraints on the superfield $\{ \boldsymbol{q},\bar{\boldsymbol{q}} \}$ \cite{BIKL1}.
Unfortunately, as it often happened  in the coset approach,  the direct covariantization of the irreducibility constraints
is not covariant~\cite{BG2}, while the simultaneous covariantization of the constraints and the equations of motion works perfectly.
That is why we propose the following equations which should describe our superparticle
\be\label{N16-dynamics}
\nabla^{i a} \mpsi_{j b} = 0, \qquad \overline{\nabla}_{i a} \bar \mpsi^{j b} =0.
\ee
These equations are covariant with respect to both unbroken and broken supersymmetries.
One should wonder whether the equations \p{N16-dynamics} are self-consistent? Indeed, due to
eqs. \p{N16-kin} from \p{N16-dynamics} we have
\be\label{selfcon}
\nabla^{i a} \mpsi_{j b} =\frac{1}{2} \nabla^{ia} \nabla_{jb} \boldsymbol{q} =0 \qquad
\Rightarrow \qquad \{ \nabla^{ia},\nabla_{jb}\}\, \boldsymbol{q}=0.
\ee
So, one may expect some additional conditions on the superfield $\boldsymbol{q}$ due to the relations \p{d1N16covder2}.
However, on the constraints surface in \p{N16-dynamics} we have
\be\label{selfcon1}
\left\{ \nabla^{ia} , \nabla^{jb} \right\}=0, \qquad
\left\{ \overline{\nabla}_{ia} , \overline{\nabla}_{jb} \right\}= 0,
\ee
and thus the equations \p{N16-dynamics} are perfectly self-consistent.

It is worth mentioning that the rest of the commutators in \p{d1N16covder2} are also simplified, when \p{N16-dynamics}
are satisfied. Indeed, on the constraints \p{N16-dynamics} surface they read
\be\label{CDalgebra}
\left \{\nabla^{i a} , \overline{\nabla}_{j b} \right \} = - 2i \delta^i_j \delta^a_b (1+ \mlambda \bar \mlambda)\nabla_t, \quad
\left [\nabla_t, \nabla^{i a} \right ] = 2i \bar \mlambda \nabla_t \mpsi^{i a} \nabla_t,\quad
\left [ \nabla_t, \overline{\nabla}_{i a} \right ] = 2i \mlambda \nabla_t \bar{\mpsi}_{i a} \nabla_t,
\ee
where we introduced the superfields $\{ \mlambda, \bar\mlambda \}$
\be\label{lambda}
\left\{
\begin{array}{l}
\overline{\nabla}_{i a} \mpsi_{j b} + \eps_{ij} \eps_{a b} \mlambda =0,\\
\nabla^{i a} \bar \mpsi^{j b} + \eps^{ij} \eps^{a b} \bar \mlambda =0,
\end{array}\right.
\quad \Rightarrow \quad
\left\{
\begin{array}{l}
\nabla_t \boldsymbol{q} +\frac{i \mlambda}{1+ \mlambda \bar \mlambda} = 0\,, \\
\nabla_t \bar{\boldsymbol{q}} - \frac{i \bar \mlambda}{1+ \mlambda \bar \mlambda} = 0.
\end{array}\right.
\ee

The superfield equations \p{N16-dynamics} lead in the bosonic limit to the following equation
of motion for the complex scalar field $q= \boldsymbol{q}|_{\theta=0}$:
\be\label{beom}
\frac{d}{dt}\left[\frac{\dot{q}}{\sqrt{1- 4 \dot{q} \dot{\bar q}}}\right]=0.
\ee
The last equation can be easy deduced from the bosonic action
\be\label{bS}
S_{bos}= \int dt \left( 1 - \sqrt{1-4 \dot{q} \dot{\bar q}}\right).
\ee
Thus, the bosonic action for a particle in $D=3$ space-time is known.

\subsection{Superparticle in D=3: component action}
Despite the explicit construction of the proper equations of motion within the superfield version of the coset approach,
it is poorly adapted for the construction of the action. That is why in the paper \cite{BKS1} the component version of
the coset  approach to construct the actions has been proposed. In the application to the present case, the basic steps
of this method can be formulated as follows:
\begin{itemize}
\item Firstly, on-shell our $N=8$ supermultiplet $\{ \boldsymbol{q}, \bar{\boldsymbol{q}} \}$ contains the following physical components:
$$
q= \boldsymbol{q}|_{\theta=0}, \quad \bar q= \bar{\boldsymbol{q}}|_{\theta=0}, \quad
\psi_{ia}=\mpsi_{ia}|_{\theta=0}, \quad \bpsi^{ia}= \bar\mpsi^{ia}|_{\theta=0}.
$$
They are just the first components of the superfields parameterizing the coset \p{cosetN16}.
\item Secondly, with respect to broken supersymmetry $\delta\theta=\delta\bar\theta=0$ \p{N16tran-B}. This means,
that the transformation properties of the physical components $\{q,{\bar q}, \psi_{ia}, \bpsi^{ia}\}$ under broken
supersymmetry can be extracted from the coset
\be\label{cosetVA}
g|_{\theta=0}=e^{i tP}\, e^{i (q Z+\bar{q} \overline Z)}\,
e^{\psi_{i a} S^{i a} + \bar{\psi}^{i a} \overline S_{i a}}.
\ee
In other words, the fields $\{q,{\bar q}, \psi_{ia}, \bpsi^{ia}\}$ parameterize the coset \p{cosetVA} which is responsible
for full breaking of the $S$ supersymmetry. Moreover, with respect to this supersymmetry the fields $\{q, {\bar q}\}$ are
just ``matter fields'', because $\delta_S q= \delta_S {\bar q}=0$, while the fermions $\{\psi_{ia}, \bpsi^{ia} \}$ are
just Goldstone fermions. This means that the component action has to be of the Volkov-Akulov type \cite{VA}, i.e. the
fermions $\{ \psi_{ia}, \bpsi^{ia} \}$ may enter the action through the einbein $\cal E$ or through
the covariant derivatives ${\cal D}_t q, {\cal D}_t{\bar q}$ only, with
\be\label{cov1}
\partial_t={\cal E}{\cal D}_t, \quad {\cal E} = E|_{\theta=0}=1- i \left( \psi_{ia} \dot{\bar\psi}{}^{ia}
+ \bar \psi^{ia} \dot \psi_{ia}  \right), \quad
{\cal E}^{-1} = 1+ i \left( \psi_{ia} {\cal D}_t \bar \psi^{ia} + \bar \psi^{ia} {\cal D}_t \psi_{ia}  \right).
\ee

Thus, the unique candidate to be the component on-shell action, invariant with respect to spontaneously broken $S$ supersymmetry  reads
\be\label{action1}
S= \alpha \int dt + \int dt {\cal E} {\cal F}\left[ {\cal D}_t q {\cal D}_t{\bar q}\right]
\ee
with an arbitrary, for the time being, function $\cal F$ and a constant parameter $\alpha$.
\item Finally, considering the bosonic limit of the action \p{action1} and comparing it with the known
bosonic action \p{bS} one may find the function ${\cal F}$:
\be\label{F}
\int dt \left( \alpha +  {\cal F}\left[\dot q \dot{\bar q}\right]\right) =\int dt\left(1-\sqrt{1-4 \dot{q} \dot{\bar q}}\right)\;
\Rightarrow\; {\cal F} =\left(1-\alpha -\sqrt{1-4 \dot{q} \dot{\bar q}}\right)\;.
\ee
\end{itemize}
Therefore, the most general component action possessing the proper bosonic limit \p{bS}  and invariant under
spontaneously broken supersymmetry has the form
\be\label{action2}
S= \alpha \int dt + (1-\alpha) \int dt\, {\cal E} -\int dt\, {\cal E} \sqrt{1- 4 {\cal D}_t q {\cal D}_t{\bar q}}\;.
\ee
In principle, the invariance of the action \p{action2} under broken supersymmetry is evident. Nevertheless, it should be explicitly checked.

From \p{N16tran-B} we obtain the total variations of our components and the time coordinate $t$:
\be\label{invB1}
\delta_S t = i \left(\eta_{i a} \bar \psi^{i a} + \bar \eta^{i a} \psi_{i a}  \right), \quad
\delta_S\psi_{i a} = \eta_{i a}, \quad \delta_S\bar \psi^{i a} = \bar \eta^{i a}, \quad
\delta_S q =0, \quad \delta_S \bar q = 0.
\ee
Therefore, the transformations of the components in the fixed point read
\be\label{inB2}
\delta^*_S q = \delta_S q - \delta_S t\, \dot q , \quad \delta^*_S \psi_{ia} = \delta_S \psi_{ia} - \delta_S t\, \dot{\psi_{ia}}.
\ee
Then, it immediately follows from \p{inB2} and definitions \p{cov1} that
\be\label{invB3}
\delta^*_S\left( {\cal E} {\cal F}\left[ {\cal D}_t q {\cal D}_t{\bar q}\right]\right)=
-i \partial_t \left[ \left( \eta_{i a} \bar \psi^{i a} + \bar \eta^{i a} \psi_{i a}  \right) {\cal E} {\cal F}
\left[ {\cal D}_t q {\cal D}_t{\bar q}\right]\right]\;.
\ee
Thus, the two last terms in the action \p{action2} are invariant, while the invariance of the first term is evident.

The final step is to check the invariance of the action \p{action2} under unbroken supersymmetry which is
realized on the components as follows:
\bea\label{Ususy1}
&& \delta^*_Q q = -2 \eps^{i a}  \psi_{i a} + i \left( \eps^{i a} \psi_{i a}\bar\lambda+
\bar\eps^{i a}\bpsi_{i a} \lambda\right) \partial_t q \nn \\
&& \delta^*_Q \psi_{i a} = \bar\eps_{i a}  \lambda + i \left( \eps^{j b} \psi_{j b}\bar\lambda+
\bar\eps^{j b}\bpsi_{j b} \lambda\right) \partial_t \psi_{i a} \;.
\eea
Here, $\lambda$ is the first component of the superfield $\mlambda$ defined in \p{lambda}
\be
\lambda = \frac{2 i {\cal D}_t q}{1+\sqrt{1- 4 {\cal D}_t q {\cal D}_t{\bar q}}}\;.
\ee
From \p{Ususy1} and the definitions \p{cov1} one may easily find the transformation properties of the main ingredients
\bea\label{Ususy2}
&& \delta^*_Q {\cal E} = i \partial_t \left[ \left( \eps^{j b}\psi_{j b}  \bar\lambda + \bar\eps^{j b}\bpsi_{j b}
\lambda \right) {\cal E}   \right] -2 i \left( \eps^{j b}\dot\psi_{j b}  \bar\lambda + \bar\eps^{j b}\dot\bpsi_{j b}  \lambda  \right), \nn \\
&& \delta^*_Q {\cal D}_t q = i \left(  \eps^{j b}\psi_{j b}  \bar\lambda + \bar\eps^{j b}\bpsi_{j b}  \lambda   \right)
\partial_t ({\cal D}_t q) -2\eps^{j b}{\cal D}_t \psi_{j b} + 2 i \left( \eps^{j b}{\cal D}_t\psi_{j b}
\bar\lambda + \bar\eps^{j b}{\cal D}_t\bpsi_{j b}  \lambda  \right){\cal D}_t q.
\eea
Now, one may calculate the variation of the integrand in the action \p{action1}
\bea\label{Ususy3}
&&\delta^*_Q \left({\cal  E}\; {\cal F}  \right) = 2\partial_t \left[ {\cal E}\, \frac{\eps^{j b}\psi_{j b} {\cal D}_t \bar q
- \bar\eps^{j b}\bpsi_{j b}  {\cal D}_t q }{1+\sqrt{1-4{\cal D}_t q  {\cal D}_t \bar q}}\, {\cal F} \right]+ \nn \\
&&+ \frac{\eps^{j b}\dot\psi_{j b} {\cal D}_t \bar q - \bar\eps^{j b}\dot\bpsi_{j b}
{\cal D}_t q}{1+\sqrt{1-4{\cal D}_t q {\cal D}_t \bar q}} \left[ -4 {\cal F} -2{\cal F}^\prime \left( 1+\sqrt{1-4{\cal D}_t q {\cal D}_t \bar q}
- 4{\cal D}_t q {\cal D}_t \bar q \right)  \right].
\eea
Substituting the function ${\cal F}$ \p{F} and its derivative over its argument ${\cal D}_t q {\cal D}_t \bar q$, we find that
the second term in the variation \p{Ususy3} cancels out, provided $\alpha=2$.
Keeping in mind that the first term in the action \p{action2} is trivially invariant under unbroken supersymmetry,
we conclude that the unique component action invariant under both unbroken and broken $N=8$ supersymmetries reads
\be\label{action3}
S= 2 \int dt -  \int dt\, {\cal E} \left(1 + \sqrt{1- 4 {\cal D}_t q {\cal D}_t{\bar q}}\right)\;.
\ee

We end this section with two comments.

\noindent
Firstly, one should note that the construction of the component action, we considered in the previous section, has two interesting peculiarities:
\begin{itemize}
\item It is based on the coset realization of the $N=16$ superalgebra \p{N16algebra}
\item In the component action \p{action3} the summation over indices $\{i,a\}$ of two $SU(2)$ groups affected
only physical fermions $\{ \psi_{ia}, \bpsi^{ia} \}$.
\end{itemize}
It is quite clear, that in such a situation one may consider two subalgebras of $N=16$ superalgebra:
\begin{itemize}
\item $N=8$ supersymmetry, by choosing the corresponding supercharges as
\be\label{N8} {\tilde Q}^{i} \equiv Q^{i1},\quad {\widetilde\bQ}_{i} \equiv \bQ_{i1}, \quad
{\tilde S}^{i} \equiv S^{i2}, \quad {\widetilde\bS}_{i} \equiv \bS_{i2},
\ee
\item $N=4$ supersymmetry with the supercharges
\be\label{N4} {\hat Q} \equiv Q^{11},\quad {\widehat\bQ} \equiv \bQ_{11}, \quad {\hat S}\equiv S^{22}, \quad {\widehat\bS} \equiv \bS_{22}.
\ee
\end{itemize}
It is evident that the corresponding component actions will be given by the same expression \p{action3}, in which
the ``new'' einbeins and covariant derivatives read
\be\label{covN8}\mbox{ $N=8$ case: } \left\{
\partial_t=\tilde{\cal E}\tilde{\cal D}_t, \quad \tilde{\cal E} = 1- i \left( \psi_{i2} \dot{\bar\psi}{}^{i2}
+ \bar \psi^{i2} \dot \psi_{i2}  \right), \quad
\tilde{\cal E}^{-1} = 1+ i \left( \psi_{i2} \tilde{\cal D}_t \bar \psi^{i2} + \bar \psi^{i2} \tilde{\cal D}_t \psi_{i2}  \right),
\right.
\ee
\be\label{covN4}\mbox{ $N=4$ case: } \left\{
\partial_t=\hat{\cal E}\hat{\cal D}_t, \quad \hat{\cal E} = 1- i \left( \psi_{22} \dot{\bar\psi}{}^{22}
+ \bar \psi^{22} \dot \psi_{22}  \right), \quad
\hat{\cal E}^{-1} = 1+ i \left( \psi_{22} \hat{\cal D}_t \bar \psi^{22} + \bar \psi^{22} \hat{\cal D}_t \psi_{22}  \right).
\right.
\ee
Thus, we see that the action \p{action3} has a universal character, describing the series of theories with
the following patterns of global supersymmetry breaking
$N=16\rightarrow N=8$, $N=8\rightarrow N=4$ and $N=4\rightarrow N=2$.

\noindent
Secondly, it is almost evident, that the universality of the action \p{action3} can be used to extend our construction
to the cases of $N=4\cdot 2^k$ supersymmetries by adding the needed numbers of $SU(2)$ indices to the superscharges as
\be\label{genSUSY1}
Q \rightarrow Q^{\alpha_1\ldots\alpha_k},\; \bQ \rightarrow \bQ_{\alpha_1\ldots\alpha_k},\quad S \rightarrow S^{\alpha_1\ldots\alpha_k},\; \bS \rightarrow \bS_{\alpha_1\ldots\alpha_k},
\ee
obeying the $N=4\cdot 2^k$ Poincar\'e superalgebra
\bea\label{genalgebra}
&& \left\{ Q^{\alpha_1\ldots\alpha_k}, \overline Q_{\beta_1\ldots\beta_k}  \right\}=2\delta^{\alpha_1}_{\beta_1}\ldots \delta^{\alpha_k}_{\beta_k} P\,, \quad
\left\{ S^{\alpha_1\ldots\alpha_k}, \overline S_{\beta_1\ldots\beta_k}  \right\}=2\delta^{\alpha_1}_{\beta_1}\ldots \delta^{\alpha_k}_{\beta_k} P\,,\nn \\
&&\left\{ Q^{\alpha_1\ldots\alpha_k}, S^{\beta_1\ldots\beta_k}  \right\}=2 i \eps^{\alpha_1 \beta_1}\ldots \eps^{\alpha_k \beta_k} Z\,, \quad
\left\{ \overline Q_{\alpha_1\ldots\alpha_k}, \overline S_{\beta_1\ldots\beta_k}  \right\}=-2 i \eps_{\alpha_1 \beta_1}\ldots \eps_{\alpha_k\beta_k} \overline Z \,,.
\eea
Once again, the component action describing superparticles in $D=3$ space with $N=4\cdot 2^k$ Poincar\'e supersymmetry
partially broken down to the $N=2\cdot 2^k$ one will be given by the same expression \p{action3} with the following substitutions
\be\label{genpsi}
\psi \rightarrow \psi_{\alpha_1\ldots\alpha_k},\; \bpsi \rightarrow \bpsi^{\alpha_1\ldots\alpha_k},\;
{\cal E} = 1- i \left( \psi_{\alpha_1\ldots\alpha_k} \dot{\bar\psi}{}^{\alpha_1\ldots\alpha_k}
+ \bar \psi^{\alpha_1\ldots\alpha_k} \dot \psi_{\alpha_1\ldots\alpha_k}  \right).
\ee

\setcounter{equation}{0}
\section{Superparticle in D=5}
In this section we will apply our approach to $N=16$ superparticle in $D=5$. The corresponding superfield equations
of motion for this system, which possesses 8 manifest and 8 spontaneously broken supersymmetries, have been constructed
within the coset approach in~\cite{BIK2}, while the action is still unknown.

To describe the superparticle in $D=5$ with 16 supersymmetries one has to start with the following superalgebra
\be\label{algebra4Z}
\{Q^i_{\alpha}, Q^j_{\beta} \} = \eps^{ij} \Omega_{\alpha \beta} P, \quad
\{Q^i_{\alpha}, S^{b \beta} \} = \delta^{\beta}_{\alpha} Z^{ib}, \quad
\{S^{a \alpha}, S^{b \beta} \} = - \eps^{ab} \Omega^{\alpha \beta} P,\quad (i,a=1,2;\, \alpha,\beta =1,2,3,4)
\ee
where the invariant $Spin(5)$ symplectic metric $\Omega_{\alpha\beta}$, allowing to raise and lower the spinor indices,
obeys the conditions\footnote{We use the following convention: $\eps^{\alpha \beta \lambda \sigma} \eps_{\alpha \beta \lambda \sigma} = 24\,,\;
\eps^{\alpha \beta \lambda \sigma}\eps_{\alpha \beta \mu \rho} = 2 (\delta^{\lambda}_{\mu}\, \delta^{\sigma}_{\rho} -
\delta^{\lambda}_{\rho}\, \delta^{\sigma}_{\mu})$.}
\bea\label{Omega}
&&
\Omega_{\alpha \beta} = - \Omega_{\beta \alpha}\,, \quad
\Omega^{\alpha \beta} = - \frac{1}{2}\, \eps^{\alpha \beta \lambda \sigma} \Omega_{\lambda \sigma}\,, \quad
\Omega_{\alpha \beta} = - \frac{1}{2}\, \eps_{\alpha \beta \lambda \sigma} \Omega^{\lambda \sigma}\,, \quad
\Omega_{\alpha \beta} \Omega^{\beta \gamma} = \delta_{\alpha}^{\gamma}\,.
\eea
From the one-dimensional perspective this algebra is $N=16$ super Poincar\'{e} algebra with four central charges $Z^{ia}$.
If we are going to treat $S$ supersymmetry to be spontaneously broken, then we have to consider the following element
of the coset:\footnote{Here, we strictly follow the notations adopted in \cite{BIK2} which are slightly different with
those we used in the previous sections.}
\be\label{coset16}
g = e^{tP}\, e^{\theta_i^{\alpha} Q^i_{\alpha}}\, e^{\boldsymbol{q}_{ia} Z^{ia}}\,
e^{\mpsi_{a \alpha} S^{a \alpha}}.
\ee
Here $(t, \theta_i^{\alpha})$ are the coordinates of $N=8, d=1$ superspace while $\boldsymbol{q}_{ia} = \boldsymbol{q}_{ia}(t,\theta_i^{\alpha}),$
$\mpsi_{a \alpha}= \mpsi_{a \alpha}(t,\theta_i^{\alpha}),$  are the Goldstone superfields.

Similarly to the case considered in the previous section, one may
find the transformation properties of the coordinates and
superfields, by acting from the left on the coset element
\p{coset16} by different elements of the group with constant
parameters. So, for the unbroken supersymmetry
($g_0=\exp{(\eps^{\alpha}_i Q_{\alpha}^i)}$) one gets
\be\label{U-susy} \delta_Q t = - \frac{1}{2}\,\eps^{\alpha}_i
\theta^{i \beta} \Omega_{\alpha \beta}\,, \quad \delta_Q
\theta^{\alpha}_i = \eps^{\alpha}_i\,, \ee while for the broken
supersymmetry ($g_0=\exp{(\eta_{a \alpha} S^{a \alpha})}$) the
corresponding transformations read \be\label{B-susy} \delta_S t =
- \frac{1}{2}\, \eta^a_{\alpha} \mpsi_{a \beta} \Omega^{\alpha
\beta}\,, \quad \delta_S \mpsi_{a \alpha} = \eta_{a \alpha}\,,
\quad \delta_S \boldsymbol{q}_{ia} = - \eta_{a
\alpha}\theta^{\alpha}_i\,. \ee The last needed ingredient is the
Cartan forms, defined in a standard way as \be\label{full-Cartan}
g^{-1} dg = \omega_P P + (\omega_Q)_i^{\alpha}\, Q^i_{\alpha} +
(\omega_Z)_{ia}\, Z^{ia} + (\omega_S)_{a \alpha}\, S^{a \alpha}\,,
\ee with \bea\label{Cartan} && \omega_P= dt - \frac{1}{2}\,d
\theta_i^{\alpha} \theta^{i \beta} \Omega_{\alpha \beta} +
\frac{1}{2}\,
 d \mpsi_{a \alpha} \mpsi^a_{\beta} \Omega^{\alpha \beta},\quad
 (\omega_Z)_{ia} = d \boldsymbol{q}_{ia} -  d \theta^{\alpha}_i \mpsi_{a \alpha} \nn \\
&& (\omega_Q)_i^{\alpha} =  d \theta^{\alpha}_i, \quad
(\omega_S)_{a \alpha} = d \mpsi_{a \alpha}.
\eea
Using the covariant differentials $\{\omega_P,(\omega_Q)_i^{\alpha}\}$ one may construct the covariant derivatives
$\nabla_t$ and $\nabla^i_{\alpha}$
\bea\label{cov-der}
\partial_t &=& E\, \nabla_t\,, \quad
E = 1 + \frac{1}{2}\, \Omega^{\beta \gamma} \mpsi^a_{\beta} \partial_t \mpsi_{a \gamma}\,, \quad
E^{-1} = 1 - \frac{1}{2}\, \Omega^{\beta \gamma} \mpsi^a_{\beta} \nabla_t \mpsi_{a \gamma}\,,\\
\nabla^i_{\alpha} &=& D^i_{\alpha} + \frac{1}{2}\, \Omega^{\beta \gamma} \mpsi^a_{\beta}
D^i_{\alpha} \mpsi_{a \gamma} \nabla_t
= D^i_{\alpha} + \frac{1}{2}\, \Omega^{\beta \gamma} \mpsi^a_{\beta}
\nabla^i_{\alpha} \mpsi_{a \gamma} \partial_t\,,
\eea
where
\be\label{flat-der1}
D^i_{\alpha} = \frac{\partial}{\partial \theta^{\alpha}_i}
+\frac{1}{2}\, \theta^{i \beta} \Omega_{\alpha \beta} \partial_t\,, \quad
\Big \{D^i_{\alpha}, D^j_{\beta} \Big \} = \eps^{ij}\, \Omega_{\alpha \beta}\, \partial_t\,.
\ee
These covariant derivatives satisfy the following (anti)commutation relations:
\bea\label{relations-der}
&&
\Big \{\nabla^i_{\alpha}, \nabla^j_{\beta} \Big \} = \eps^{ij}\, \Omega_{\alpha \beta}\,
\nabla_t + \Omega^{\lambda \sigma}\, \nabla^i_{\alpha} \mpsi^b_{\lambda}\,
\nabla^j_{\beta} \mpsi_{b \sigma}\, \nabla_t\,,\nn \\
&&
\Big [\nabla_t, \nabla^i_{\alpha} \Big ] = \Omega^{\beta \gamma}\, \nabla_t \mpsi^b_{\beta}\,
\nabla^i_{\alpha} \mpsi_{b \gamma} \nabla_t\,.
\eea

Now, in a full analogy with the previously considered case, we impose the following invariant condition on the
$d\theta$-projections of Cartan form $(\omega_Z)_{ia}$  \p{Cartan}:
\be
(\omega_Z)_{ia}|_\theta=0 \quad \Rightarrow \quad \left\{
\begin{array}{l}
\nabla^{(j}_{\alpha}\, \boldsymbol{q}^{i)}_a = 0\,,  \qquad \qquad \quad \quad \quad \;\quad \quad\; \mbox{(a)}\\
\nabla^i_{\alpha}\, \boldsymbol{q}_{ia} - 2 \mpsi_{a \alpha}=0.\;\;\; \qquad \qquad \quad \quad \mbox{(b)}
\end{array}\right. \label{EoM-Zt}
\ee
The condition $(\ref{EoM-Zt}b)$ identifies the fermionic superfield $\mpsi_{a\alpha}$ with the spinor derivatives of
the superfield $\boldsymbol{q}_{ia}$, just reducing the independent superfields to bosonic $\boldsymbol{q}_{ia}$ ones (this is again the Inverse Higgs
Phenomenon \cite{ih}). The conditions $(\ref{EoM-Zt}a)$ are more restrictive - they nullify all auxiliary components
in the superfield $\boldsymbol{q}_{ia}$. Indeed, it immediately follows from \p{EoM-Zt} that
\be\label{dopeq1}
\frac{3}{2} \nabla_\beta^j \mpsi_{a\alpha} = \left\{ \nabla_\beta^j, \nabla_\alpha^i\right\} \boldsymbol{q}_{ia} -
\frac{1}{2} \left\{ \nabla_\alpha^j, \nabla_\beta^i\right\}\boldsymbol{q}_{ia}.
\ee
Using anti-commutators \p{relations-der}, one may solve this equation as follows:
\be\label{EoM-Stt}
\nabla^j_{\beta}\, \mpsi_{a \alpha} + \frac{1}{2}\, \mlambda^j_a \Omega_{\alpha \beta} = 0\,,
\ee
where the superfield $\mlambda^{ia}$ is defined as $(\mlambda^2 = \mlambda^{ia} \mlambda_{ia})$
\be\label{lambdaB}
\nabla_t\, \boldsymbol{q}^{ia} - \frac{1}{2}\, \frac{\mlambda^{ia}}{1+\frac{\mlambda^2}{8}} = 0.
\ee
Thus, we have the on-shell situation. In \cite{BIK2} the corresponding bosonic equation of motion has been found to be
\be\label{boseom}
\frac{d}{dt} \left( \frac{\dot q_{ia}}{\sqrt{1-2 \dot q^{jb} \dot q_{jb}}}\right)=0,
\ee
where $q_{ia}=\boldsymbol{q}_{ia}|_{\theta=0}$ are the first components of the superfield $\boldsymbol{q}_{ia}$.
The equation of motion \p{EoM-Stt}
corresponds to the static-gauge form of Nambu-Goto action for the massive particle in $D=5$ space-time
\be\label{NG5}
S_{bos} \sim \int dt\left(1- \sqrt{1- 2 \dot q^{ia} \dot q_{ia}}\right) .
\ee

To construct the on-shell component action we will follow the same procedure which was described above in
full details. So, we will omit unessential details concentrating only on the new
features.

If we are interested in the invariance with respect to broken $S$ supersymmetry, then we may consider the reduced
coset element
\be\label{coset_red}
g|_{\theta=0} = e^{tP}\,  e^{q_{ia} Z^{ia}}\,
e^{\psi_{a \alpha} S^{a \alpha}}.
\ee
Here, $q_{ia}$ and $\psi_{a\alpha}$ are the first components of the superfields $\boldsymbol{q}_{ia}$ and
$\mpsi_{a\alpha}$. Similarly to the discussion in section 3, the Goldstone fermions $\psi_{a\alpha}$
may enter the component action only through the einbein $\cal E$ and the covariant derivatives ${\cal D}_t q_{ia}$, defined now as
\be\label{cd4c}
\partial_t = {\cal E}\, {\cal D}_t\,, \quad
{\cal E} = 1 + \frac{1}{2}\, \Omega^{\beta \gamma} \psi^a_{\beta} \partial_t \psi_{a \gamma}\,, \quad
{\cal E}^{-1} = 1 - \frac{1}{2}\, \Omega^{\beta \gamma} \psi^a_{\beta} {\cal D}_t \psi_{a \gamma}\,,
\ee
Keeping in the mind the known bosonic limit of the action \p{NG5}, we come to the unique candidate of the component on-shell action
\be\label{action24}
S= \alpha \int dt + (1-\alpha) \int dt\, {\cal E} -\int dt\, {\cal E} \sqrt{1- 2 {\cal D}_t q^{ia} {\cal D}_t q_{ia}}\;.
\ee
This action is perfectly invariant with respect to broken $S$ supersymmetry, realized on the physical components and
their derivatives as
\be\label{susy-B1}
\delta_S^* q_{ia} = \frac{1}{2}\, \eta^b_{\alpha} \psi_{b \beta} \Omega^{\alpha \beta}
\partial_t q_{ia}\,, \quad
\delta_S^* (\cD_t\,q_{ia}) = \frac{1}{2}\, \eta^b_{\alpha} \psi_{b \beta} \Omega^{\alpha \beta} \partial_t (\cD_t\,q_{ia})\,, \quad
\delta_S^* \psi_{a \alpha} = \eta_{a \alpha} + \frac{1}{2}\, \eta^b_{\beta} \psi_{b \lambda} \Omega^{\beta \lambda}
\partial_t \psi_{a \alpha}\,.
\ee
From \p{susy-B1} one may find the transformation properties of the einbein ${\cal E}$
\be\label{var-E-B}
\delta_S^* {\cal E} = \frac{1}{2}\,\eta^a_{\alpha} \partial_t \left( {\cal E} \Omega^{\alpha \beta} \psi_{a \beta}\right )\,.
\ee
Now, combining \p{susy-B1}  and \p{var-E-B}, we will get
\be\label{var1gen}
\delta_S^*\left( {\cal E} {\cal F}\left[ \cD_t\,q^{jb}\cD_t\,q_{jb}\right]\right) =
\frac{1}{2}\,\eta^a_{\alpha} \partial_t \left( \Omega^{\alpha \beta} \psi_{a \beta} \; {\cal E}\; {\cal F}\left[ \cD_t\,q^{jb}\cD_t\,q_{jb}\right]\right )\,,
\ee
and, therefore, the second and the third terms in the action \p{action24} are separately invariant with respect
to $S$ supersymmetry. The first term in \p{action24} is trivially invariant with respect to both, broken and unbroken supersymmetries.

The last step is to impose invariance with respect to unbroken $Q$ supersymmetry. Under the transformations of
unbroken supersymmetry taken in the fixed point the variation of any superfield reads
$$\delta^*_Q {\boldsymbol{F}}= - \eps^{\alpha}_i Q^i_{\alpha}{\boldsymbol{F}}\;.$$
From this one may find the variations of the components $q_{ia}$ and $\psi_{a \alpha}$
and their covariant derivatives:
\bea\label{susy-U1}
&&
\delta_Q^* q_{ia} = - \eps^{\alpha}_i \psi_{a \alpha}
+ \frac{1}{4}\, \eps^{\alpha}_j \lambda^{jb} \psi_{b \alpha} \partial_t q_{ia}\,,\nn \\
&&
\delta_Q^* (\cD_t q_{ia}) = - \eps^{\alpha}_i \cD_t \psi_{a \alpha}
+ \frac{1}{4}\, \eps^{\alpha}_j  \frac{\lambda_{ia}}{1+\frac{1}{8}\,\lambda^2}\,
\lambda^{jb}\cD_t \psi_{b \alpha}
+ \frac{1}{4}\,\eps^{\alpha}_j \lambda^{jb} \psi_{b \alpha} \partial_t (\cD_t q_{ia})\,, \nn \\
&&\label{susy-U2}
\delta_Q^* \psi_{a \alpha} = \frac{1}{2}\, \eps^{\beta}_j \Omega_{\alpha \beta} \lambda^j_a
+ \frac{1}{4}\, \eps^{\beta}_j \lambda^{jb} \psi_{b \beta} \partial_t \psi_{a \alpha}\,.
\eea
The variation of the einbein ${\cal E}$ can be also computed and it reads
\be\label{var-E-U}
\delta_Q^* {\cal E} = \frac{1}{4}\,\eps^{\beta}_j \partial_t \left ( {\cal E} \lambda^{j b} \psi_{b \beta} \right )
-  \frac{1}{2}\,\eps^{\beta}_j \lambda^{j b} \partial_t \psi_{b \beta}\,.
\ee
It is a matter of lengthy, but straightforward calculations to check that the action \p{action24} is invariant under
unbroken supersymmetry \p{susy-U2}, \p{var-E-U} if $\alpha=2$.

Thus, the component action, invariant under both unbroken and broken $N=8$ supersymmetries, reads
\be\label{N16-actionZZ}
S = \int dt \left [2-{\cal E}\, \left ( 1+ \sqrt{1-2 \cD_t q^{ia} \cD_t q_{ia}}\right )
\right ]\,.
\ee

\setcounter{equation}{0}
\section{Supermembrane in D=4}
As an instructive application of our approach we consider in this section as an example two models, namely,
the supermembrane in $D=4$ and the supersymmetric space-filling $\rm{D2}$-brane.
We will mainly follow the paper \cite{IK}.
\subsection{Supermembrane in D=4: kinematical constraints, equations of motion and the component action}
The nonlinear realization of the breaking $N=1, D=4 \rightarrow N=1,d=3$ has been constructed in \cite{IK}.
There, the $N=1, D=4$ super Poincar\'{e} group has been realized in its coset over the $d=3$ Lorentz group $SO(1,2)$
\be\label{cosetA}
g=e^{x^{ab}P_{ab}}e^{\theta^{a}Q_{a}}e^{\boldsymbol{q} Z}
e^{\mpsi^a S_a} e^{\boldsymbol{\Lambda}^{ab} K_{ab}} \;.
\ee
Here, $x^{ab}, \theta^a$ are $N=1, d=3$ superspace coordinates, while the remaining coset parameters are Goldstone superfields,
$\mpsi^a = \mpsi^a(x,\theta),\;\boldsymbol{q} = \boldsymbol{q}(x,\theta),\; \boldsymbol{\Lambda}^{ab}= \boldsymbol{\Lambda}^{ab}(x,\theta)$.
To reduce the number of independent superfields one has to impose the constraints\footnote{We collect the exact expressions
for the covariant derivatives $\nabla_{ab}, \nabla_a$ and their properties, constructed in \cite{IK}, in  Appendix A.}
\be
\Omega_Z = 0 \quad \Rightarrow \quad \left\{
\begin{array}{l}
\nabla_{ab} \boldsymbol{q} + \frac{4}{1+2\boldsymbol{\lambda}^2} \boldsymbol{\lambda}_{ab} = 0, \quad\; \mbox{(a)}\\
\nabla_a \boldsymbol{q} - \mpsi_a = 0.\;\; \qquad \qquad \mbox{(b)}
\end{array}\right. \label{ihA}
\ee
Eqs. \p{ihA} allow us to express $\boldsymbol{\lambda}_{ab}(x,\theta)$ and $\mpsi^a(x,\theta)$ through covariant derivatives of $\boldsymbol{q}(x,\theta)$.
Thus, the bosonic superfield $\boldsymbol{q}(x,\theta)$ is the only essential Goldstone superfield we need for this case
of the partial breaking of the global supersymmetry.
The constraints \p{ihA} are covariant under all symmetries and they do not imply any dynamics and leave $\boldsymbol{q}(x,\theta)$ off-shell.

The last step we can make within the coset approach is to write the covariant superfield equations of motion.
It was  shown in \cite{IK} that this can be achieved  by imposing the following constraint on the Cartan form:
\be
\Omega_S| =0 \quad \Rightarrow \quad \left\{
\begin{array}{l}
\nabla^a \mpsi_a = 0, \qquad \qquad \mbox{(a)} \\
\nabla_{(a} \mpsi_{b)}= -2\boldsymbol{\lambda}_{ab}. \;\quad \mbox{(b)}
\end{array}\right. \label{eom_form}
\ee
where $|$ denotes the ordinary $d\theta$-projection of the form $\Omega_S$.

Eqs. \p{eom_form} imply the proper dynamical equation of motion
\be\label{peom}
\nabla^a \nabla_a \boldsymbol{q}=0.
\ee
This equation is also covariant with respect to all symmetries, and its bosonic limit
for $q(x)= \boldsymbol{q}(x,\theta)|_{\theta = 0}$ reads
\be  \label{NGeq}
\partial_{ab}\left( \frac{\partial^{ab}q}
{\sqrt{1-\frac{1}{2}\partial q \cdot \partial q}} \right) =0~,
\ee
which corresponds to the ``static gauge'' form of the $D=4$
membrane Nambu-Goto action
\be  \label{NG}
S= \int d^3x \left( 1 - \sqrt{1-\frac{1}{2}\partial^{ab} q  \partial_{ab} q}
\right)\, .
\ee
Thus, the equations \p{eom_form} indeed describe the supermembrane in $D=4$.

Until now we just repeated the standard coset approach steps from the paper \cite{IK} in the application to
the $N=1, D=4$ supermembrane. As was already mentioned in section 2, the nonlinear realization approach  fails to construct
the superfield action. That is why, to construct the superfield action one has to involve some additional arguments/scheme
as it has been done, for example, in \cite{IK}.

Funny enough, if we instead will be interested in the component action, then it can be constructed almost immediately within
the nonlinear realization approach. One may check that all important features  of the on-shell (i.e. with Eqs.
\p{eom_form} taken into account) component action we summarized in section 2, are present in the case at hands.
Indeed,
\begin{itemize}
\item All physical components, i.e. $\boldsymbol{q}|_{\theta=0}$ and $\mpsi^a|_{\theta=0}$, are among the ``coordinates'' of our coset
\p{cosetA} as the $\theta=0$ parts of the corresponding superfields,
\item Under spontaneously broken supersymmetry the superspace coordinates $\theta^a$ do not transform at all \p{susy2}.
Therefore, the corresponding transformation properties of the fermionic components $\boldsymbol{\psi}^a|_{\theta=0}$ are {\it the same
as in the Volkov-Akulov model} \cite{VA}, where  all supersymmetries are supposed to be spontaneously broken,
\item Finally, the $\theta=0$ component of our essential Goldstone superfield $\boldsymbol{q}(x,\theta)$ does not transform
under spontaneously broken supersymmetry and, therefore, it behaves like a ``matter'' field within the Volkov-Akulov
scheme.
\end{itemize}
As the immediate consequences of these features we conclude that
\begin{itemize}
\item The  fermionic components $\mpsi^a|_{\theta=0}$ may enter the component action either through $\det {\cal E}$ \p{E1}
(to compensate the transformation of volume $d^3 x$ under \p{susy2}) or through the covariant derivatives ${\cal D}_{ab}$ \p{nabla}, only,
    \item The ``matter'' field $q = \boldsymbol{q}|_{\theta=0}$ may enter the action only through covariant derivatives ${\cal D}_{ab} q$.
\end{itemize}

Thus, the unique candidate to be the component on-shell action, invariant with respect to spontaneously broken supersymmetry $S$ reads
\be\label{cact1}
S= \alpha \int d^3x +\beta \int d^3x\, \det {\cal E} {\cal F}({\cal D}^{ab} q\, {\cal D}_{ab}q),
\ee
with an arbitrary, for the time being, function $\cal F$. All other interactions between the bosonic component $q$ and
the fermions of spontaneously broken supersymmetry $\psi^a$ are forbidden!

Note, that the first, trivial term in \p{cact1} is independently invariant under broken (and unbroken!) supersymmetries, because,
in virtue of \p{susy2}
\be
\delta_S \int d^3x \sim \int d^3x \, \partial_{ab} \left( \xi^a \psi^b\right)\quad \mbox{and, therefore }\quad
\delta_S \int d^3x=0.
\ee
As we already said in section 2, this term in the action \p{cact1} ensures the validity of the  limit
$S_{q=0,\psi=0} =0$.

The action \p{cact1} is the most general component action invariant with respect to broken supersymmetry.
But in the present case we explicitly know its bosonic limit - it should be just the Nambu-Goto action \p{NG}.
Some additional
information about its structure comes from the linearized form of the action, which,  according with its invariance
with respect to unbroken supersymmetry, has to be
\be
S_{lin}\sim \psi^a \partial_{ab} \psi^b - \frac{1}{4} \partial^{ab}q \partial_{ab}q.
\ee
Combining all these ingredients, which completely fix the parameters $\alpha$ and $\beta$ in \p{cact1}, we can write
the component action of $N=1, D=4$ supermembrane as
\be\label{Action}
S= \int d^3 x\left[ \;2 - \det {\cal E}\;\left(1+\sqrt{1-\frac{1}{2} {\cal D}^{ab} q {\cal D}_{ab} q} \right)\right].
\ee
The explicit expression for $\det {\cal E}$ has the form
\bea\label{detE}
\det {\cal E} &=& 1+\frac{1}{2} \psi^a {\cal D}_{ab} \psi^b-\frac{1}{16} \psi^d\psi_d\; {\cal D}^{ab} \psi^c {\cal D}_{ab}\psi_c =\nn \\
&=&1+\frac{1}{2} \psi^a \partial_{ab} \psi^b+\frac{1}{8}\psi^d\psi_d \left( \partial^{ab}\psi_b \partial_{ac}\psi^c+
\frac{1}{2} \partial^{ab}\psi^c \partial_{ab}\psi_c\right).
\eea

Let us stress, that such a simple form of the component action is achieved only in the rather specific basis, where the
bosonic $q$ and fermionic fields $\psi^a$ are the Goldstone fields of the nonlinear realization. Surely, this choice is not
unique and in  different bases  the explicit form of  action could drastically change. The most illustrative example
is given by the action in~\cite{AGIT}, where the on-shell component action for the supermembrane  has been constructed for the first time.

The detailed proof that the action \p{Action} is invariant with respect to both, broken and unbroken supersymmetries,
can be found in Appendix B.

\subsection{Supersymmetric  space-filling D2-brane}
Due to the duality between scalar field and gauge field strength in $d=3$, the action for ${\rm D2}$-brane can be easily constructed
within the coset approach. The idea of the construction is similar to the purely bosonic case. The crucial step is to treat
the first, bosonic component of $\boldsymbol{\lambda}_{ab}$ as an independent component (i.e. to ignore the (a) part of Eqs.\p{ihA}).
Now, the generalized variant of the action \p{Action} reads
\be\label{d2act1}
S=\int d^3x \left[ 2- \det {\cal E} -\det {\cal E} \left( 1+ 2 \frac{\lambda^{ab}({\cal D}_{ab}q +2 \lambda_{ab})}{1-2 \lambda^2}\right)\right].
\ee
All these summands have a description in terms of $\theta=0$ parts of the Cartan forms \p{cartan}.
The first term is just a volume form constructed from ordinary differentials $dx^{ab}$. The second terms is a volume form constructed from
semi-covariant differentials $d{\hat x}{}^{ab}$
$$ d{\hat x}^{ab}  =   dx^{ab}+
  \frac{1}{4}\psi^a d\psi^b +\frac{1}{4}\psi^{b}d\psi^{a}.$$
Finally, the last term in \p{d2act1} is a volume form constructed from the $\theta=0$ component of the form $\Omega_P^{ab}$ \p{cartan}
$$ d{\tilde x}^{ab} = d{\hat x}^{ab}+ \frac{2}{1-2\lambda^2} \lambda^{ab}\left( {\cal D}_{cd} q +2 \lambda_{cd}\right)d{\hat x}^{cd}. $$

Since the action \p{d2act1} depends only on $\lambda^{ab}$ and not on its derivatives, the $\lambda$-equation of motion
\be\label{D2eom1}
{\cal D}_{ab} q =- \frac{4\lambda_{ab}}{1+2 \lambda^2}
\ee
can be used to eliminate $\lambda^{ab}$ in favor of ${\cal D}_{ab} q$. Clearly, the Eq. \p{D2eom1} is just the
 $(a)$ part of the constraints \p{ihA}, we ignored while introducing the action \p{d2act1}. Plugging $\lambda_{ab}$ expressed
 through ${\cal D}_{ab} q$ back into \p{d2act1} gives us the action \p{Action}.

Alternatively, the equation of motion for $q$
\be\label{D2eom2}
\partial_{ab} \left[ \frac{ \det {\cal E}\; \lambda^{cd}\; \left( {\cal E}^{-1}\right)_{cd}{}^{ab}}{1-2\lambda^2}\right]=0
\ee
has the form of the $d=3$ Bianchi identity for the field strength $F^{ab}$
\be\label{D2FS}
F^{ab}\equiv \frac{ \det {\cal E}\; \lambda^{cd}\; \left( {\cal E}^{-1}\right)_{cd}{}^{ab}}{1-2\lambda^2} \qquad \Rightarrow \qquad
\partial_{ab} F^{ab}=0.
\ee
Substituting this into the action \p{d2act1} and integrating by parts, one may bring it to the supersymmetric ${\rm D2}$-brane action
\be\label{BI}
S=\int d^3x \left[ 2 -\det {\cal E}\left( 1+\sqrt{1+ 8 {\widetilde F}{}^2}\right)\right]\,,
\ee
where
\be
{\widetilde F}_{ab} \equiv \frac{{\cal E}_{ab}{}^{cd}\;F_{cd}}{\det {\cal E}} = \frac{\lambda_{ab}}{1-2\lambda^2}\,.
\ee
Therefore,
\be\label{BII}
S=2 \int d^3 x \left[ 1- \det{\cal E} \;\frac{1}{1-2 \lambda^2}\right].
\ee
Clearly, in the bosonic limit ${\widetilde F}_{ab}=F_{ab}$ and thus, the bosonic part of the \p{BI} is the standard
Born-Infeld action for ${\rm D2}$-brane, as it should be.

\setcounter{equation}{0}
\section{Supermembrane in  D=5}
In this section we construct the on-shell component action for $N=1, D=5$ supermembrane
and its dual versions, corresponding to a vector and a double vector supermultiplets.
We demonstrate that the proper choice of the components and using the covariant (with respect to broken supersymmetry) derivatives drastically
simplify the action: it can be represented as the sum of four terms each having an explicit geometric meaning.
\subsection{Supermembrane}
In the present case we are dealing with spontaneous breaking of $N=1, D=5$ Poincar\'{e} supersymmetry down to $N=2, d=3$ one.
From the $d=3$standpoint the $N=1, D=5$ supersymmetry algebra is a central-charges extended $N=4$ Poincar\'{e} superalgebra
with the following basic anticommutation relations:
\be\label{basicAL}
\left\{ Q_{a} , \bQ_{b}  \right\} =2P_{ab}, \quad \left\{ S_a, \bS_b \right\}=2P_{ab},  \quad \left\{ Q_{a}, S_{b} \right\}
=2\epsilon_{ab} Z, \quad \left\{ \bQ_{a}, \bS_{b} \right\} =2\epsilon_{ab} \bZ.
\ee
The $d=3$ translations generator $P_{ab}$ and the central charge generators $Z, \bZ$ form $D=5$ translation generators.
We will also split the generators of $D=5$ Lorentz algebra $so(1,4)$ into $d=3$ Lorentz algebra generators $M_{ab}$,
the generators $K_{ab}$ and $\bK_{ab}$ belonging to the coset $SO(1,4)/SO(1,2)\times U(1)$ and $U(1)$ generator $J$.
The full set of (anti)commutation relations can be found in the Appendix C.

Keeping $d=3$ Lorentz and, commuting with it, $U(1)$ subgroups of $D=5$ Lorentz group $SO(1,4)$ linearly realized, we will
choose the coset element as
\be\label{a_coset}
g = e^{i x^{ab}P_{ab}}e^{\theta^a Q_a + \bar\theta^a \bQ_a}e^{i (\boldsymbol{q} Z+\bar{\boldsymbol{q}} \bZ)}e^{\mpsi^a S_a + \mbpsi^a \bS_a}
e^{i  (\boldsymbol{\Lambda}^{ab}K_{ab}+ \bar{\boldsymbol{\Lambda}}^{ab}\bK_{ab})}.
\ee
Here, $\left\{x^{ab}, \theta^a, \bar\theta^a\right\}$ are $N=2, d=3$ superspace coordinates, while the remaining coset
parameters are $N=2$ Goldstone superfields. The whole $N=1, D=5$ Poincar\'{e} supergroup can be realized in this coset
by the left acting on \p{a_coset} of the different elements of the supergroup. We summarize in Appendix C the resulting transformation properties of
the coordinates and superfields with respect to unbroken \p{susy1A}, broken \p{susy2A} supersymmetries and automorphism \p{auto},
as well as a pure technical calculation of Cartan forms, semi-covariant derivatives and their superalgebra
\p{Dq}, \p{nabla1}, \p{deralg}.

Similarly to the previously considered cases, to reduce the number of independent
superfields one has to impose the constraints
\be\label{IH}
\Omega_Z =0 \quad \Rightarrow \quad \left\{
\begin{array}{l}
\nabla_{ab} \boldsymbol{q} =- 2 i\, \frac{(1+\ml \cdot \mbl) \ml_{ab} -\ml^2 {\mbl}_{ab}}{(1+\ml\cdot\mbl)^2-\ml^2{\mbl}{}^2}\,,\\
\nabla_a \boldsymbol{q} = -2 i\, \mpsi_a,\quad \overline \nabla_a \boldsymbol{q} =0,
\end{array} \right. \quad
{\overline\Omega}_Z =0 \quad \Rightarrow \quad \left\{
\begin{array}{l}
\nabla_{ab} \bar{\boldsymbol{q}} = 2 i\, \frac{(1+\ml \cdot \mbl) \mbl_{ab} -\mbl^2 \ml_{ab}}{(1+\ml\cdot\mbl)^2-\ml^2{\mbl^2}},\\
\overline \nabla_a \bar{\boldsymbol{q}} = -2 i\, \bar\mpsi_a,\quad \nabla_a \bar{\boldsymbol{q}} =0.
\end{array} \right.
\ee
Here, in order to simplify the expressions, we have passed to the some variant of the stereographic parametrization of the
coset $SO(1,4)/SO(1,2)\times U(1)$
\be\label{l}
\ml_{ab}=\left( \frac{\tanh\sqrt{\boldsymbol{Y}}}{\sqrt{\boldsymbol{Y}}}\right)_{ab}^{cd}\;\mLambda_{cd},\quad
\mbl_{ab}=\left( \frac{\tanh\sqrt{\boldsymbol{Y}}}{\sqrt{\boldsymbol{Y}}}\right)_{ab}^{cd}\;\mbLambda_{cd}.
\ee
The equations \p{IH} allow us to express superfields $\mLambda_{ab}, \mbLambda_{ab}$ and $\mpsi^a, \mbpsi^a$ through covariant derivatives
of $\boldsymbol{q}(x,\theta,\bar\theta)$ and $\bar{\boldsymbol{q}}(x,\theta,\bar\theta)$. Thus, the bosonic superfields $\boldsymbol{q}(x,\theta,\bar\theta),
\bar{\boldsymbol{q}}(x,\theta,\bar\theta)$ are the only essential Goldstone superfields  needed for this case of the partial breaking of the global supersymmetry.
The constraints \p{IH} are covariant under all symmetries, they do not imply any dynamics and leave $\boldsymbol{q}(x,\theta,\bar\theta)$
and $\bar{\boldsymbol{q}}(x,\theta,\bar\theta)$ off-shell.

Within the coset approach we may also write the covariant superfield equations of motion. This can be achieved by imposing the proper constraint on the
 Cartan forms for broken supersymmetry. In the present case these constraints read
\bea\label{eomCF}
&& \left. \Omega_S \right|=0\; \Rightarrow \; (a)\; \nabla_a \mpsi_b = 0, \quad (b)\; \bar\nabla_b \mpsi^a=- i\, \mLambda_b{}^c
\left( \frac{\tan 2 \sqrt{ \overline{\boldsymbol{T}}}}{ \sqrt{ \overline{\boldsymbol{T}}}}\right)_c^a \equiv
- i\, \mlambda_b^a \nn \\
&& \left. {\overline\Omega}_S \right|=0\; \Rightarrow \; (a)\; \bar\nabla_a \mbpsi_b = 0, \quad (b)\; \nabla_b \mbpsi^a= i\, \mLambda_b{}^c
\left( \frac{\tan 2 \sqrt{ {\boldsymbol{T}}}}{ \sqrt{ {\boldsymbol{T}}}}\right)_c^a \equiv i, \mblambda_b^a,
\eea
where $|$ means the $d\theta$-projection of the forms.

Let us make a few comments concerning the constraints given above:
\begin{itemize}
\item The easiest way to check that the equations \p{IH}, \p{eomCF} put the theory on-shell  is to consider these equations in the linearized form
    \bea
    && \partial_{ab} \boldsymbol{q} =-2 i\, \mLambda_{ab} \; (a), \quad D_a \boldsymbol{q} =-2 i\, \mpsi_a\; (b),
    \quad \bD_a \boldsymbol{q}=0 \; (c),\label{lin1} \\
    && D_a \mpsi_b =0 \; (a), \quad \bD_b \mpsi^a =- 2 i\, \mLambda_b^a\; (b). \label{lin2}
    \eea
    Acting on eq. (\ref{lin1}b) by $\bD_b$ and using the eq.(\ref{lin1}c) and the algebra of spinor derivatives \p{flatCD} we
    immediately conclude that eq. (\ref{lin2}b) follows from \p{lin1}. In addition, the eq. (\ref{lin2}a)
    means that the auxiliary component of the superfield $\boldsymbol{q}$ is zero and, therefore, our system is on-shell
    \be\label{lin3}
    D_a \mpsi_b =0\; \Rightarrow \; D^2 \boldsymbol{q} =0\; \Rightarrow \partial_{ab} D^b \boldsymbol{q} =0\;\Rightarrow \Box \boldsymbol{q}=0.
    \ee
\item It turns out that the variables $\{\mlambda_a^b, \mblambda_a^b\}$ defined in \p{eomCF}, are more suitable than the
$\{\ml_{ab}, \mbl_{ab}\}$ \p{l} one. Using the algebra of covariant derivatives \p{deralg} it is easy to find the following
relations from \p{IH} and \p{eomCF}:
\be\label{qlambda}
\nabla_{ab} \boldsymbol{q} =- i\, \frac{ \mlambda_{ab}-\frac{1}{2} \mlambda^2 \mblambda_{ab}}
{1-\frac{1}{4} \mlambda^2 \mblambda^2},\quad \nabla_{ab}\bar{\boldsymbol{q}} = i\, \frac{ \mblambda_{ab}-\frac{1}{2}
\mblambda^2 \mlambda_{ab}}{1-\frac{1}{4} \mlambda^2 \mblambda^2}.
\ee
These equations play the same role as those in \p{IH}, relating the superfields $\{\mlambda_{ab}, \mblambda_{ab}\}$
(and, therefore, the superfields $\{\mLambda_{ab}, \mbLambda_{ab}\}$) with the space-time derivatives of the superfields
$\{\boldsymbol{q}, \bar{\boldsymbol{q}}\}$.
\end{itemize}

Now we present two different ways to construct the bosonic action.\\
The first of them is based on the consideration of the bosonic coset related to \p{a_coset}
and on the invariance of constraints \p{IH}, \p{eomCF} with respect to all $N=1, D=5$ Poincar\'{e} supergroup.
Thus we have
\be\label{bos_coset}
g_{bos} = e^{i\, x^{ab}P_{ab}}e^{i\,( q Z+\bar{q} \bZ)} e^{i\,  (\Lambda^{ab}K_{ab}+ \bLambda^{ab}\bK_{ab})}.
\ee
Clearly, the corresponding bosonic Cartan forms can be easily extracted from their superfields version \p{Dx}. The bosonic
version of the constraints \p{IH} results in the relations
\be\label{boIH}
\partial_{ab} q =- 2i\, \frac{(1+l \cdot \bar{l}) l_{ab} -l^2 {\bar l}_{ab}}{(1+l\cdot\bar{l})^2-l^2{\bar l}{}^2}\,,\quad
\partial_{ab}{\bar q} = 2i\, \frac{(1+l \cdot \bar{l}) \bar{l}_{ab} -{\bar l}^2 l_{ab}}{(1+l\cdot\bar{l})^2-l^2{\bar{l}^2}}\,,
\ee
while the bosonic vielbein $\mathcal{B}_{ab}{}^{cd} = \mathcal{E}_{ab}{}^{cd}|_{\psi=0}$
\be\label{bos_e_def}
\left(\Omega_P^{bos}\right) = dx^{ab} \mathcal{B}_{ab}{}^{cd} P_{cd}
\ee
acquires the form
\be
 \mathcal{B}_{ab}^{cd}=\delta_{a}^{(c} \delta_{b}^{d)}-\frac{2}{(1+l\cdot \bar{l})^2-l^2\,
 \bar{l}^2}\left[(1+l\cdot \bar{l})\,\left(\bar{l}^{cd} l_{ab}+l^{cd} \bar{l}_{ab}\right)- \bar{l}^2\,
 l^{cd} l_{ab}- l^2\, \bar{l}^{cd} \bar{l}_{ab}\right]. \nn
\ee
Therefore, the simplest invariant bosonic action reads
\be\label{bos_action1}
S_{bos}=\int d^3x \det \mathcal{B} =
 \int d^3x \frac{(1-l\cdot \bar{l})^2-l^2\, \bar{l}^2}{(1+l\cdot \bar{l})^2-l^2\, \bar{l}^2}\,,
\ee
or in terms of $\{q, {\bar q}\}$
\be\label{bos_action2}
S_{bos}=\int d^3x \sqrt{\left(1-\partial_{ab}q \, \partial^{ab}\bar q\right)^2-\left(\partial_{ab}q\,
\partial^{ab}q\right)\,\left(\partial_{cd}\bar q \,\partial^{cd}\bar q\right)}\,.
\ee
The latter is the static gauge Nambu-Goto action for membrane in $D=5$.
One can also add the following action, trivially invariant under the transformations $ISO(1,4)$
\be\label{bosS0}
S_0 = \int d^3 x.
\ee

Another way to derive the bosonic action is to use automorphism transformation laws.
These laws \p{auto} in the bosonic limit have the form
\be\label{d3N2aut1}
\delta x^{ab} = 2i\, \left( \bar a ^{ab} q - a^{ab} \bar q   \right), \quad \delta q =-2i\, (ax), \quad \delta\bar q =2i\, (\bar a x).
\ee
The active form of these transformations reads
\be\label{d3N2aut2}
\delta^* q = -2i\, (ax)-2i\, \partial_{ab} q \left(   \bar a ^{ab} q - a^{ab} \bar q  \right), \quad
\delta^* \bar q = 2i\, (\bar ax)-2i\, \partial_{ab} \bar q \left(   \bar a ^{ab} q - a^{ab} \bar q  \right).
\ee
Due to translations, $U(1)$-rotations  and $d=3$ Lorentz invariance, the action may depend only on
the following scalar combination of partial derivatives of bosons $\{q, \bar q\}$
\be
\xi = \partial_{ab}q \;\partial^{ab}\bar q,\quad \eta = \partial_{ab}q\; \partial^{ab}q,\quad \bar \eta= \partial_{ab}{\bar q}\; \partial^{ab}{\bar q}\,,
\ee
which in accordance  with \p{d3N2aut2} transforms as
\bea\label{d3N2aut3}
\delta^* \xi &=& 2 i\, (\bar a \partial q) -2 i\, (a\partial \bar q) -2 i\, (\bar a^{ab}q
- a^{ab}\bar q)\partial_{ab}\xi -2 i\, (\bar a \partial q)\xi + 2 i\, (a \partial \bar q)\xi
-2 i\, (\bar a \partial \bar q)\eta + 2 i\, (a\partial q)\bar\eta\,,\nn  \\
\delta^* (\eta\bar\eta) &=& 4 i\, (\bar a \partial \bar q)\eta -4 i\, (a\partial q)\bar\eta
-2 i\, (\bar a^{kl} q - a^{kl} \bar q)\partial_{kl}(\eta\bar\eta)-4 i\, (\bar a \partial q) \eta\bar\eta
+ 4 i\, (a\partial \bar q)\eta\bar\eta + \nn \\
&& 4 i\, (a\partial q)\xi\bar\eta -4 i\, (\bar a \partial \bar q)\xi\eta\,.
\eea
Therefore, the variation of the arbitrary function ${\cal F}(\xi,\eta\bar\eta)$ reads
\bea\label{d3N2aut4}
\delta^* {\cal F} &=& 2i\, \left[ (a\partial q)\bar\eta - (\bar a \partial \bar q) \eta \right]
\left( {\cal F}_\xi + 2(\xi-1){\cal F}_{(\eta\bar\eta)}   \right) + 2i\, \left[ (\bar a \partial q) - (a\partial \bar q)
\right]\left({\cal F}+ (1-\xi){\cal F}_\xi -2\eta\bar\eta {\cal F}_{(\eta\bar\eta)}  \right)
\nn \\
&&
- 2i\,\partial_{ab}\left[ \left(q {\bar a}^{ab} - {\bar q} a^{ab}\right){\cal F} \right].
\eea
Thus, to achieve the invariance of the action one has impose the following restrictions on the function ${\cal F}$:
\be\label{d3N2aut5}
{\cal F}_\xi +2(\xi -1){\cal F}_{(\eta\bar\eta)} =0, \quad {\cal F} + {\cal F}_\xi (1-\xi) -2(\eta\bar\eta) {\cal F}_{(\eta\bar\eta)} =0,
\ee
with the evident solution
\be
{\cal F} = \sqrt{(1-\xi)^2 -\eta\bar\eta}\,.
\ee
Therefore, the invariant action has the form
$$
S = \int d^3 x\, \sqrt{(1-\partial_{ab} q \partial^{ab}\bar q)^2 -(\partial_{ab} q \partial^{ab} q)(\partial_{kl}\bar q \partial^{kl}\bar q)}\,,
$$
and thus, it coincides with that previously constructed in \p{bos_action2}, as it should be.

Let us now construct the full component action for supermembrane which will be invariant under both broken and unbroken supersymmetries.
We begin our analysis with the broken supersymmetry $S$.\\
The superspace coordinates $\{\theta,\bar\theta\}$ of the coset \p{a_coset} do not transform under
$S$ supersymmetry. Therefore, each component of superfields transforms independently under the broken supersymmetry.
Thus, from \p{susy2A} one finds
\be\label{Str}
\delta x^{ab}= i\, \left(\varepsilon^{(a}\bpsi^{b)}+\bar\varepsilon^{(a}\psi^{b)}\right),\quad
\delta q=0,\quad \delta \bar{q}=0,\quad \delta\psi^a=\varepsilon^a,\quad \delta\bpsi^a=\bar\varepsilon^a.
\ee
Then, one may easily check that the $\theta=0$ projections of the covariant differential $\triangle x^{ab}$ \p{Dx}
\be\label{Sdx}
\hat{\triangle} x^{ab} \equiv \triangle x^{ab}|_{\theta=0} = dx^{ab} - i\, \left(
\psi^{(a} d\bpsi^{b)} + \bpsi^{(a} d\psi^{b)}\right)\equiv {\cal E}^{ab}_{cd}\; dx^{cd}\,,
\ee
as well as the covariant derivatives constructed from them
\be\label{SD}
\cD_{ab} = \left( {\cal E}^{-1} \right)_{ab}^{cd}\; \partial_{cd}
\ee
are also invariant under broken supersymmetry. From all this it immediately follows that the action possessing the proper
bosonic limit \p{bos_action2} and invariant under broken supersymmetry reads
\be\label{Saction}
S_1 = \int d^3 x\, \det {\cal E} \sqrt{(1-\cD_{ab}\, q \cD^{ab}\bar q)^2 -(\cD_{ab}\, q \cD^{ab} q)(\cD_{cd}\bar q\, \cD^{cd}\bar q)}\;.
\ee
The action \p{Saction} reproduces the kinetic terms for the bosonic and fermionic components
\be\label{Saction_lin}
S_1 = \int d^3 x\, \left[ -i\, \left( \psi^a \partial_{ab} \bpsi^b + \bpsi^a \partial_{ab} \psi^b \right) -
\partial_{ab}q \partial^{ab}{\bar q} + \ldots\right],
\ee
but the coefficient between them is strictly fixed. This could be not enough to maintain unbroken supersymmetry.
So, one has to add to \p{Saction} the purely fermionic action
\be\label{Saction2}
S_2 =\int d^3 x\, \det {\cal E}\,,
\ee
which is trivially invariant under broken supersymmetry. Finally, in order to have a proper limit
$$
S_{q\rightarrow 0, \psi \rightarrow 0} =0,
$$
one has to involve into the game the trivial action $S_0$ that reads as
\be\label{Saction0}
S_0 = \int d^3 x.
\ee
Thus, the Ansatz for the supersymmetric action acquires the form
\bea\label{Saction_gen}
S&=& \left( 1 + \alpha\right) S_0 - S_1 - \alpha S_2 \nn \\
&=&\left( 1 + \alpha\right)\int d^3 x -
\int d^3 x\, \det {\cal E} \left( \alpha+  \sqrt{(1-\cD_{ab} q \cD^{ab}\bar q)^2 -(\cD_{ab} q \cD^{ab} q)(\cD_{cd}\bar q \cD^{cd}\bar q)}\right)\,,
\eea
where the constant $\alpha$ has to be defined.\\
In the previously considered cases in the above sections, the Ansatz, similar to \p{Saction_gen},
was completely enough to maintain the second, unbroken supersymmetry. A careful analysis shows that in the present case
there is an additional Wess-Zumino term which has to be taken into account
\be\label{WZ}
S_{WZ}= i\,  \int d^3 x \, \det {\cal E} \;\left( \psi^m \cD_{ab} \bpsi_m - \bpsi^m \cD_{ab} \psi_m \right) \cD^{ac}\; q\; \cD_c{}^b \; {\bar q}\,.
\ee
The variation of $S_{WZ}$ under $S$ supersymmetry reads (note, that only the variations of $\psi, \bpsi$ without derivatives play a role)
\be\label{inv1}
\delta S_{WZ} =i\,  \int d^3 x \,\det {\cal E} \;\left( \varepsilon^m \cD_{ab} \bpsi_m - \bar\varepsilon^m \cD_{ab} \psi_m \right)
\cD^{ac} q\; \cD_c{}^b  {\bar q}\,.
\ee
The simplest way to check that $\delta S_{WZ}=0$ is to pass to the $d=3$ vector notations
\footnote{Our conventions to pass to/from vector indices are summarized in Appendix C, \p{44}.}.
Then we have
\bea\label{inv2}
\delta S_{WZ} &\sim &  \int d^3 x \;\det {\cal E} \;\epsilon^{IJK}\left( \varepsilon^m \cD_I \bpsi_m
- \bar\varepsilon^m \cD_I \psi_m \right) \cD_J\; q\; \cD_K \; {\bar q}  \nn \\
&\sim& \int d^3 x \;\det {\cal E} \;\det {\cal E}^{-1} \epsilon^{IJK}\left( \varepsilon^m \partial_I \bpsi_m
- \bar\varepsilon^m \partial_I \psi_m \right) \partial_J\; q\; \partial_K \; {\bar q} \nn \\
&\sim& \int d^3 x \;\partial_I\left[ \epsilon^{IJK}\left( \varepsilon^m \bpsi_m
- \bar\varepsilon^m \psi_m \right) \partial_J\; q\; \partial_K \; {\bar q}\right] =0.
\eea
Thus, the action $S_{WZ}$ \p{WZ} is invariant under $S$ supersymmetry and our Ansatz for
the membrane action is extended to be
\be\label{ACTION}
S= \left( 1 + \alpha\right) S_0 - S_1 - \alpha S_2+\beta S_{WZ}.
\ee
Thus, after imposing broken supersymmetry, the component action \p{ACTION} is fixed up
to two constants $\alpha$ and $\beta$. No other terms or structures are admissible!

Now we are going to demonstrate how the unbroken supersymmetry fixes these constants.
In order to maintain the unbroken supersymmetry, one has to find the transformation properties of all
objects presented in \p{ACTION}.
Using the transformations of the superspace coordinates \p{susy1A} one gets
for the $\epsilon$-part of the transformations
\bea
&&
\delta\psi_a = -\epsilon^b \left.\left(D_b \mpsi_a\right)\right|_{\theta=0} = \epsilon^b \psi^m\blambda_b^n \partial_{mn}\psi_a\,,\nn\\
&&
\delta\cD_{ab}\psi_c =-\epsilon^d \left.\left( D_d \nabla_{ab}\mpsi_c\right)\right|_{\theta=0} =
2 \epsilon^d \cD_{ab}\psi^m \blambda_d^n \cD_{mn}\psi_b+\epsilon^d \psi^m\blambda_d^n\partial_{mn}\cD_{ab}\psi_c\,, \nn\\
&&
\delta\cD_{ab}q = -\epsilon^d \left.\left( D_d \nabla_{ab} \boldsymbol{q} \right)\right|_{\theta=0} =
2\epsilon^d \cD_{ab}\psi^m \blambda_d^n\cD_{mn} q + 2 i\, \epsilon^d \cD_{ab} \psi_d+
\epsilon^d\psi^m \blambda_d^n \partial_{mn}\cD_{ab} q\,, \label{Qsusytr}
\eea
and, as the consequence,
\be\label{SEtr}
\delta \det {\cal E} = \partial_{mn}\left[ \epsilon^d \psi^m \blambda_d{}^n \det {\cal E}\right] -
2 \epsilon^d \blambda_d{}^n \cD_{mn}\psi^n \det {\cal E}\,.
\ee
In order to fix the parameter $\alpha$ one may consider just the kinetic terms in the action \p{ACTION}
\be\label{Skin}
S_{kin} = \int d^3 x\, \left[ -i\, \left(\alpha+1\right) \left( \psi^a \partial_{ab} \bpsi^b
+ \bpsi^a \partial_{ab}\psi^b\right) +\partial_{ab} q\, \partial^{ab}{\bar q}\right],
\ee
which has to be invariant under the linearized form of the transformations \p{Qsusytr}
(see also \p{lin1}, \p{lin2})
\be\label{linQ}
\delta \bpsi_a =- i\, \epsilon^b\blambda_{ba}\simeq-\epsilon^b \partial_{ba}{\bar q}, \quad \delta\partial_{ab}q =
2i\, \epsilon^d \partial_{ab}\psi_d.
\ee
Varying the integrand in \p{Skin} and integrating by parts, we get
\be
\delta S_{kin} = \int d^3 x\, \left[ 2i\, (\alpha+1) \epsilon^c \psi^a\partial_{ab} \partial_c{}^b {\bar q}
- 2i\, \epsilon^d \psi_d \Box {\bar q} \right]
= \int d^3 x\, \left[ i\, (\alpha+1) \epsilon^d \psi_d \Box {\bar q} - 2i\, \epsilon^d \psi_d \Box {\bar q} \right].
\ee
Therefore, we have to fix the constant $\alpha$ as
\be\label{alpha}
\alpha=1.
\ee
The fixing of the last parameter $\beta$ is more involved.
Using the transformation properties \p{Qsusytr} one may find
\be
\delta \cF = 2\left( \epsilon^c \blambda_c^n \cD_{ab} \psi^m \cD_{nm} q
+i\, \epsilon^c \cD_{ab}\psi_c\right)\, \frac{\partial \cF}{\partial \cD_{ab}q}+ 2 \epsilon^c \blambda_c^n \cD_{ab}\psi^m \cD_{mn}{\bar q}\,
\frac{\partial \cF}{\partial \cD_{ab}{\bar q}}
+ \epsilon^c \blambda_c^n \psi^m \partial_{mn} \cF,
\ee
where
\be\label{FF}
\cF \equiv \sqrt{(1-\cD_{ab} q\, \cD^{ab}\bar q)^2 -(\cD_{ab} q\, \cD^{ab} q)(\cD_{cd}\bar q\, \cD^{cd}\bar q)}\;.
\ee
In order to avoid the appearance of the square roots, it proved to be more convenient to use the equalities
\be
\frac{\partial \cF}{\partial \cD_{ab}q} =- i\, \frac{\blambda^{ab}+\frac{1}{2} \blambda^2 \lambda^{ab}}{1-\frac{1}{4}\lambda^2 \blambda^2}\,, \quad
\frac{\partial \cF}{\partial \cD_{ab}{\bar q}} = i\, \frac{\lambda^{ab}+\frac{1}{2} \lambda^2 \blambda^{ab}}{1-\frac{1}{4}\lambda^2 \blambda^2}\,.
\ee
Performing a straightforward calculation  one gets
\be
\delta \left[- \det {\cal E} \left(1+\cF\right)\right]= 2 i\, \epsilon^c \det {\cal E} \left( \cD_{ab}\psi_c\, \cD^{ab}{\bar q} -
2\cD_{am}\psi^m\, \cD_c^a {\bar q}\right) -
2 \epsilon^c \det {\cal E} \blambda_{cm} \cD_{ab}\psi^m\,  \cD^{ad}q\, \cD_d^b {\bar q}\,.
\label{deltaF}
\ee
Similarly, one may find the variation of the integrand of the action $S_{WZ}$ (up to the surface terms disappearing after integration over $d^3x$)
\be
\delta {\cal L}_{WZ}= -2 \beta \epsilon^c \det {\cal E} \left[ \left( \psi^k \cD_{ab}\bpsi_k -\bpsi^k \cD_{ab}\psi_k\right) \cD^{ad}\psi_c\, \cD_d^b {\bar q}
- \blambda_{cm} \cD_{ab}\psi^m\, \cD^{ad} q\, \cD_d^b{\bar q}\,\right]. \label{deltaWZ}
\ee
Now, it is a matter of quite lengthly, but again straightforward calculations, to check that the sum of
variations \p{deltaF} and \p{deltaWZ} is a surface term if
\be\label{beta}
\beta=1 .
\ee
Thus, we conclude that the action of the supermembrane in $D=5$, which is invariant with respect to both unbroken
and broken supersymmetries, has the form
\bea
S &=& 2\int d^3 x -
\int d^3 x\, \det {\cal E} \left( 1+  \sqrt{(1-\cD_{ab} q\, \cD^{ab}\bar q)^2 -(\cD_{ab} q\, \cD^{ab} q)(\cD_{cd}\bar q\, \cD^{cd}\bar q)}\right) \nn \\
&+& i\,  \int d^3 x \, \det {\cal E} \left( \psi^m \cD_{ab} \bpsi_m - \bpsi^m \cD_{ab} \psi_m \right) \cD^{ac} q\, \cD_c{}^b {\bar q}\,. \label{finAction}
\eea

\subsection{Dualization of the scalars: vector and double vector supermultiplets}
Due to the duality between scalar field, entering the action with the space-time derivatives only, and gauge
field strength in $d=3$, the actions for the vector (one scalar dualized) and the double vector (both scalars dualized)
supermultiplets can be easily obtained within the coset approach. Before performing such dualizations, let us
rewrite the action \p{finAction} in the vector notations. If we introduce the quantity
\be\label{G}
\cG_{ab} =\frac{1}{\sqrt{2}}\left( \psi^m \cD_{ab} \bpsi_m - \bpsi^m \cD_{ab} \psi_m\right),
\ee
then only vector indices show up in the action. Passing to the vector notation, we get
\bea\label{finAction_vec}
S &=& 2\int d^3 x-
\int d^3 x\, \det {\cal E} \left( 1+  \sqrt{(1-\cD_I q \cD_I\bar q)^2 -(\cD_{I} q \cD_I q)(\cD_{J}\bar q \cD_{J}\bar q)}\right) \nn \\
&+& i\,  \int d^3 x \,\det {\cal E} \epsilon^{IJK} \cG_I \cD_{J} q\, \cD_K {\bar q}\,,
\eea
where
\be
\cD_I = \left( {\cal E}^{-1}\right)_I{}^J \partial_J, \quad {\cal E}_I{}^J= \delta_I^J-\frac{1}{\sqrt{2}}\left(\sigma^J\right)_{ab}
\left(\psi^a \partial_I \bpsi^b +\bpsi^a \partial_I \psi^b \right).
\ee
\subsubsection{Vector supermultiplet}
The vector $N=2, d=3$ supermultiplet includes one scalar and one gauge fields
among the physical bosonic components. Thus, we have to dualize one of the scalar components in the action \p{finAction_vec}.
To perform dualization, one has to pass to a pair of real bosons $\{u, v\}$
\be
q=\frac{1}{2}(u+i v), \quad {\bar q} = \frac{1}{2}(u-i v).
\ee
In terms of the newly defined scalars, the action \p{finAction_vec} reads
\bea\label{daction1}
S &=& 2\int d^3 x-
\int d^3 x\, \det {\cal E} \left[ 1+  \sqrt{\left(1-\frac{1}{2}\cD_I u \cD_I u\right)\left(1-\frac{1}{2} \cD_J v \cD_J v\right)
-\frac{1}{4} \left(\cD_{I} u \cD_I v\right)^2 }\;\right] \nn \\
&+& \frac{1}{2}  \int d^3 x \,\det {\cal E} \;\epsilon^{IJK} \cG_I \cD_{J} u\; \cD_K  v\,.
\eea
The equation of motion for the bosonic field $v$ has the form
\be\label{deq1}
\partial_I \left( \det {\cal E} \left({\cal E}^{-1}\right)_J^I\, V_J\right)=0,\quad
V_I={\widetilde V}_I+\frac{1}{2}\epsilon_{IJK}G_J \cD_K u,
\ee
where
\be\label{deq2}
{\widetilde V}_I =\frac{\left(1-\frac{1}{2}\cD u \cdot \cD u\right) \cD_I v +\frac{1}{2}\ \cD u \cdot \cD v\; \cD_I u}
{2 \sqrt{\left(1-\frac{1}{2}\cD u \cdot \cD u\right)\left(1- \frac{1}{2}\cD v \cdot \cD v\right) -\frac{1}{4} \left(\cD u \cdot \cD v\right)^2 }}.
\ee
Then, one may find that
\be
\cD_I v = \frac{2 {\widetilde V}_I -{\widetilde V}\cdot \cD u \; \cD_I u}{\sqrt{1-\frac{1}{2} \cD u \cdot \cD u
+2 {\widetilde V}\cdot {\widetilde V} -\left( {\widetilde V}\cdot \cD u\right)^2}} .
\ee
Performing the Rauth transformation over the bosonic field $v$, we finally get
\be\label{daction11}
{\tilde S}=2 \int d^3 x- \int d^3 x\, \det {\cal E} \left( 1+\sqrt{1-\frac{1}{2} \cD u \cdot \cD u +2 {\widetilde V}\cdot {\widetilde V}
-\left( {\widetilde V}\cdot \cD u\right)^2}\;\right).
\ee
This is the action for the $N=2,d=3$ vector supermultiplet which possesses an additional, spontaneously broken $N=2$ supersymmetry.

One should stress that the real field strength is defined in \p{deq1}, but the action has a much simpler structure written
in terms of ${\widetilde V}_I$.

\subsubsection{Double vector supermultiplet}
In order to obtain a double vector supermultiplet, one may dualize both scalars in the action \p{finAction_vec}.
As the first step, one has to find the equations of motion for the scalar fields
\be\label{eomdv}
\partial_I \left( \det {\cal E} \left( {\cal E}^{-1}\right)^I_J\; V^J \right) =0, \quad
\partial_I \left( \det {\cal E} \left( {\cal E}^{-1}\right)^I_J\; \bV^J \right) =0,
\ee
where
\be\label{VV}
V_I ={\widetilde V}_I - i\, \epsilon_{IJK} G_J \cD_k {\bar q}, \quad {\widetilde V}_I=
\frac{\left(1-\cD q \cdot \cD {\bar q}\right) \cD_I {\bar q}+\left( \cD {\bar q} \cdot \cD {\bar q}\right) \cD_I q}
{\sqrt{(1-\cD q \cdot \cD \bar q)^2 -(\cD q \cdot \cD q)(\cD \bar q \cdot \cD\bar q)}}\;.
\ee
After a standard machinery with the Rauth transformations we finally get the action
\be\label{dvector}
{\widehat S}= 2\int d^3 x -\int d^3 x\, \det {\cal E} \left[1+ \sqrt{ \left( 1+ {\widetilde V}\cdot {\overline{\widetilde V}}\right)^2
-{\widetilde V}{}^2\; {\overline{\widetilde V}}{}^2} -
i\, \epsilon_{IJK}\; G_I\; {\widetilde V}_J {\overline{\widetilde V}}_K\right] .
\ee
The bosonic sector of this action coincides with that constructed in \cite{IKL22}. Again, the simplest form of the
action is achieved with the help of ${\widetilde V}_I$ variables which are related with field strengths as in \p{eomdv}, \p{VV}.

\section{Conclusion}
In this paper, using a remarkable connection between the partial
breaking of global supersymmetry, the coset approach, which realized
the specific pattern of supersymmetry breaking, and the Nambu-Goto
actions for the extended object, we have reviewed the construction of the on-shell
component actions for the superparticle in $D=3$ realizing $N=4\cdot 2^{k} \rightarrow N=2\cdot 2^k$ pattern of supersymmetry breaking, for the
superparticle in $D=5$ with the $N=16$ supersymmetry broken down to $N=8$ one,
for the $N=1, D=5$ supermembrane and its dual cousins, and for the $N=1$ supermembrane in $D=4$.
Of course, such  actions can be obtained by dimensional
reduction from the higher dimension actions or even from the known superspace actions.
Nevertheless, if we pay more attention to the spontaneously broken
supersymmetry and, thus, use the corresponding covariant
derivatives, together with the proper choice of the components,
the resulting action can be drastically simplified. So, the
implications of our results are threefold:
\begin{itemize}
\item we demonstrated that the coset approach can be used  far beyond the construction of the superfield equations of motion,
if we are interested in the component actions,
\item we showed that there is a rather specific choice of the superfields and their components which drastically simplifies the component action,
\item we argued that the broken supersymmetry fixed the on-shell component action up to some constants, while the role
of the unbroken supersymmetry is just to fix these constants.
\end{itemize}
The application of our approach is not limited to the cases of P-branes only.
Different types of D-branes could be also considered in a similar
manner. However, once we are dealing with the field strengths,
which never show up as the coordinates of the coset space, the
proper choice of the components becomes very important. In
particular, the Born-Infeld-Nambu-Goto action \p{daction1}, we
constructed by the dualization of one scalar field, has a nice,
compact form in terms of the ``covariant'' field strength
${\widetilde V}_I$ which is related with the ``genuine'' field
strength, obeying the Bianchi identity, in a rather complicated
way \p{deq1}. The same is also true for the Born-Infeld type
action \p{dvector}. In order to clarify the nature of these variables, one
has to consider the corresponding patterns of the supersymmetry
breaking (with one, or without central charges in the $N=4,d=3$
Poincar\'{e} superalgebra \p{algebra}) independently. In this
respect, the detailed analysis of $N=2 \rightarrow N=1$
supersymmetry breaking in $d=4$ seems to be much more interesting,
being a preliminary step to the construction of $N=4$ Born-Infeld
action \cite{Kallosh, BIK3, BIK4} and/or to the action describing
partial breaking of $N=1,D=10$ supersymmetry with the
hypermultiplet as the Goldstone superfield.

In this paper we also showed that the on-shell component actions for superparticle have the universal form
$$
S= \alpha \int dt + (1-\alpha) \int dt\, {\cal E} - \int dt\, {\cal E}  \sqrt{1- \beta {\cal D}_t q {\cal D}_t{q}}\;.
$$
With our approach, we explicitly constructed such actions for the superparticles in $D=3$ realizing $N=4\cdot 2^{k} \rightarrow N=2\cdot 2^k$
pattern of supersymmetry breaking, and in $D=5$ with the $N=16$ supersymmetry broken down to $N=8$ one.
It was shown that the corresponding component on-shell actions are invariant under both unbroken and broken supersymmetry.
In the considered models only the equality of both unbroken and broken supersymmetries was essential, and their number did not
play any role, we expect that all superparticle models with one half partial breaking of global supersymmetry can be constructed
similarly, confirming, thereby, its universality.

One possible application of this method is the construction of models with partial breaking of global supersymmetry
in cases when $d>2$, where the superspace actions are known (see, e.g.,~\cite{BG1,BG2,BG3,RT}).
We assume that these actions derived with our method will have a more simple
and understandable form.

It would be quite instructive to understand which new features will appear when we will replace the trivial, flat
target space by, for example, the AdS one~\cite{BIKaDs}.
It seems that the strategy will be the same, and we are planning to report
the corresponding results elsewhere.

\section*{Acknowledgements}
We thank our co-authors Nikolay Kozyrev and Armen Yeranyan with whom some of the results mentioned here were
obtained. We wish to acknowledge useful discussions with them, as well as with  S.~Kuzenko and D.~Sorokin.
This work was partially supported by RFBR
grants~12-02-00517-a, 13-02-91330-NNIO-a and 13-02-90602 Apm-a, as
well as by the ERC Advanced Grant no. 226455
\textit{``Supersymmetry, Quantum Gravity and Gauge
Fields''}~(\textit{SUPER\-FIELDS}).

\setcounter{equation}{0}
\def\theequation{A.\arabic{equation}}
\section*{Appendix A. Superalgebra, coset space, transformations and Cartan forms}
In this appendix we collected some formulas from the paper \cite{IK} where the nonlinear realization
of $N=1, D=4$ Poincar\'{e} group in its coset over $d=3$ Lorentz group $SO(1,2)$ was constructed.

In $d=3$ notation the $N=1, D=4$  Poincar\'e superalgebra contains the following set of generators:
\be\label{1}
\mbox{ N=2, d=3 SUSY }\quad \propto \quad \left\{ Q_a, P_{ab}, S_a, Z, M_{ab}, K_{ab} \right\},
\ee
$a,b=1,2$ being the $d=3$ $SL(2,R)$
spinor indices \footnote{The indices are raised
and lowered as follows:
$V^{a}=\epsilon^{ab}V_b,\;V_{b}=\epsilon_{bc}V^c,\;
\epsilon_{ab}\epsilon^{bc}=\delta_a^c\; .$}.
Here, $P_{ab}$ and $Z$ are $D=4$ translation generators,  $Q_a$ and $S_a$
are the generators of super-translations, the generators $M_{ab}$ form $d=3$ Lorentz algebra $so(1,2)$, while the generators
$K_{ab}$ belong to the coset $SO(1,3)/SO(1,2)$. The basic anticommutation relations read
\be
\left\{ Q_{a},Q_{b}\right\}=P_{ab}\; ,\quad
\left\{ Q_{a},S_{b}\right\} = \epsilon_{ab}Z\; , \quad
\left\{ S_{a},S_{b}\right\} = P_{ab} \;. \label{susy}
\ee

The coset element was defined in \cite{IK} as
\be\label{coset1Z}
g=e^{x^{ab}P_{ab}}e^{\theta^{a}Q_{a}}e^{\boldsymbol{q}Z}
  e^{\mpsi^aS_a}e^{\mLambda^{ab}K_{ab}} \,.
\ee
Here, $x^{ab}, \theta^a$ are $N=1, d=3$ superspace coordinates, while the remaining coset parameters are Goldstone superfields,
$\boldsymbol{q} = \boldsymbol{q}(x,\theta),\; \mpsi^a \equiv \mpsi^a(x,\theta),\;\mLambda^{ab} = \mLambda^{ab}(x,\theta)$.

The transformation properties of the coordinates and superfields with respect
to all symmetries can be found by acting from the left on the coset element \p{coset1Z} by the different elements
of $N=1, D=4$ supergroup. They have the following explicit form:
\begin{itemize}
\item Translations and Unbroken supersymmetry $(g_0=\mbox{exp }(a^{ab}P_{ab}+
  \eta^{a}Q_{a} ))$
\be\label{susy1}
\delta x^{ab}=a^{ab}-\frac{1}{4}\eta^a\theta^b-\frac{1}{4}\eta^b\theta^a ,
\quad
\delta \theta^{a}=\eta^a\,.
\ee
\item Broken supersymmetry $(g_0=\mbox{exp }(\xi^{a}S_{a}))$
\be\label{susy2}
\delta x^{ab}= -\frac{1}{4}\xi^a \mpsi^b-\frac{1}{4}\xi^b \mpsi^a,\quad
\delta \boldsymbol{q}=\xi^a\theta_a,\quad
\delta\mpsi^a=\xi^a \,.
\ee
\item $K$ transformations $(g_0=\mbox{exp }(r^{ab}K_{ab}))$
\bea
&&\delta x^{ab}= -2\boldsymbol{q} r^{ab}-\frac{1}{2}\theta_c r^{ca}\mpsi^b-
   \frac{1}{2}\theta_c r^{cb}\mpsi^a +
     \frac{1}{2}\theta^a r^{bc}\mpsi_c+
     \frac{1}{2}\theta^b r^{ac}\mpsi_c\,, \nn\\
&&\delta \theta^{a} = -2r^{ab}\mpsi_b \,, \quad
\delta \boldsymbol{q} = -4r_{ab}x^{ab}, \quad
\delta \mpsi^a= 2r^{ab}\theta_b,\quad
   \delta \mlambda^{ab}=r^{ab}-4 \mlambda^{ac} r_{cd} \mlambda^{db}.
   \label{ktr}
\eea
\item Broken $Z$-translations $(g_0 = \mbox{exp}(cZ))$
\be \label{Ztr}
\delta \boldsymbol{q} = c\,.
\ee
\item The $d=3$ Lorentz group $SO(1,2)\sim SL(2,R)$ acts as rotations of
the spinor indices.
\end{itemize}
In \p{ktr} the coordinates of the stereographic parametrization of
the coset $SO(1,3)/SO(1,2)$ have been defined as
\be
\mlambda^{ab}=
\frac{\tanh\left(\sqrt{2\mLambda^2}\right)}{\sqrt{2\mLambda^2}}\,
\mLambda^{ab}\,,\quad
\tanh{}^2\left(\sqrt{2\mLambda^2}\right)\equiv 2 \mlambda^2 \,,\quad
\mLambda^2 \equiv \mLambda_{ab}\mLambda^{ab}\,, \quad
\mlambda^2 \equiv \mlambda_{ab}\mlambda^{ab}.
\ee

The most important objects in the coset are the Cartan forms
$$g^{-1}d g =  \Omega_Q + \Omega_P + \Omega_Z + \Omega_S + \Omega_K +
\Omega_M .$$
In what follows we will need only the forms $\Omega_Q, \Omega_P, \Omega_Z$ and $\Omega_S$ which were constructed in \cite{IK}
\bea
\Omega_Z & = & \frac{1+2\mlambda^2}{1-2\mlambda^2}\left[ d\hat{\boldsymbol{q}} +
    \frac{4}{1+2\mlambda^2}\mlambda_{ab}d{\hat x}^{ab}\right]Z\,, \nn\\
\Omega_P &\equiv & \Omega_P^{ab}P_{ab} = \left[ d{\hat x}^{ab}+
\frac{2}{1-2\mlambda^2}
\mlambda^{ab}\left(
d\hat{\boldsymbol{q}} + 2 \mlambda_{cd}d{\hat x}^{cd}\right) \right] P_{ab} \,,\nn \\
\Omega_Q &\equiv & \Omega_Q^aQ_a = \frac{1}{\sqrt{1-2\mlambda^2}}
\Big [ d\theta^{a}+
     2\mlambda^{ab}d\mpsi_{b}\Big ]Q_{a} \,, \nn\\
 \Omega_S &\equiv & \Omega_S^a S_a = \frac{1}{\sqrt{1-2\mlambda^2}} \Big [ d\mpsi^{a}-
     2\mlambda^{ab}d\theta_{b}\Big ]S_{a} \,.
\label{cartan} \\
d{\hat x}^{ab}  &\equiv &  dx^{ab}+\frac{1}{4}\theta^{a}d\theta^{b}
  +\frac{1}{4}\theta^{b}d\theta^{a}+
  \frac{1}{4}\mpsi^a d\mpsi^b +\frac{1}{4}\mpsi^{b}d\mpsi^{a},\quad
d\hat{\boldsymbol{q}} \equiv  d\boldsymbol{q} +\mpsi_{a}d\theta^{a}. \label{dx}
\eea
Note, that all Cartan forms, except for $\Omega_M$, transform homogeneously under all symmetries.

Having at hands the Cartan forms, one may construct the ``semi-covariant'' (covariant with respect to $d=3$ Lorentz,
unbroken and broken supersymmetries only) as
\be\label{cD}
d{\hat x}^{ab}\nabla_{ab} +d\theta^a \nabla_a = dx^{ab} \frac{\partial}{\partial x^{ab}}
+ d\theta^a \frac{\partial}{\partial \theta^a}.
\ee
Explicitly, they read \cite{IK}
\be\label{nabla}
\nabla_{ab} =  (E^{-1})^{cd}_{ab}\,\partial_{cd} \,, \quad
\nabla_a = D_a + \frac{1}{2}\mpsi^b D_a \mpsi^c \,\nabla_{bc} =
D_a + \frac{1}{2}\mpsi^b \nabla_a \mpsi^c \,{\partial}_{bc}\,,
\ee
where
\bea
&&
D_a=\frac{\partial}{\partial \theta^a}+
\frac{1}{2}\theta^b\partial_{ab}\,, \quad
\left\{ D_a, D_b \right\} =\partial_{ab} \,, \label{flatd} \\
&&
E_{ab}^{cd}=\frac{1}{2}(\delta_a^c\delta_b^d+\delta_a^d\delta_b^c)+
\frac{1}{4}(\mpsi^c\partial_{ab}\mpsi^d+ \mpsi^d\partial_{ab}\mpsi^c) \,,\label{E1} \\
&&
(E^{-1})^{cd}_{ab} = \frac{1}{2}(\delta_a^c\delta_b^d+\delta_a^d\delta_b^c)-
\frac{1}{4}(\mpsi^c\nabla_{ab}\mpsi^d+ \mpsi^d\nabla_{ab}\mpsi^c) \,.
\eea
These derivatives obey the following algebra:
\bea
&& \left[ \nabla_{ab},\nabla_{cd} \right] = -\nabla_{ab}\mpsi^m \nabla_{cd}\mpsi^n \nabla_{mn} \,, \qquad
\left[ \nabla_{ab},\nabla_{c} \right] = \nabla_{ab}\mpsi^m\nabla_{c}\mpsi^n \nabla_{mn} \,, \nn \\
&& \left\{ \nabla_{a},\nabla_{b} \right\} =\nabla_{ab} + \nabla_{a}\mpsi^m\nabla_{b}\mpsi^n \nabla_{mn} \,.
\label{algebra}
\eea

\setcounter{equation}{0}
\def\theequation{B.\arabic{equation}}
\section*{Appendix B}
In this Appendix we will prove the invariance of the supermembrane action \p{Action} under broken and unbroken supersymmetries.
The proof for the broken supersymmetry is the easiest one and we will start with this invariance.
\subsection*{Broken supersymmetry}
Under spontaneously broken $S^a$ supersymmetry our coordinates and the physical components transform as in \p{susy2}
\be\label{Bsusy2}
\delta x^{ab}= -\frac{1}{4}\xi^a\psi^b-\frac{1}{4}\xi^b\psi^a,\quad
\delta q=0,\quad \delta\psi^a=\xi^a \; .
\ee
One may immediately check that the $\theta=0$ part of the covariant differential $d{\hat x}^{ab}$, defined in \p{dx}
\be\label{Bdx}
d{\hat x}^{ab}  =   dx^{ab}+
  \frac{1}{4}\psi^a d\psi^b +\frac{1}{4}\psi^{b}d\psi^{a}
\ee
is invariant under the transformations \p{Bsusy2}. Therefore, the covariant derivatives ${\cal D}_{ab} = \nabla_{ab}|_{\theta=0}$
\p{nabla} are also invariant under broken supersymmetry transformations.
Now, for the active form of the transformations
($\delta^* \phi =\phi'(x)-\phi(x)$) we have
\bea\label{Trsusy2}
&& \delta_S^* {\cal D}^{ab} q = \frac{1}{2} \xi^c \psi^d \partial_{cd} {\cal D}^{ab}q \quad \Rightarrow\quad
\delta_S^* {\cal F}({\cal D} q \cdot {\cal D} q) =\frac{1}{2} \xi^a \psi^b \partial_{ab} {\cal F}, \nn \\
&& \delta_S^* \psi^a = \xi^a+\frac{1}{2} \xi^c \psi^d \partial_{cd} \psi^a, \qquad
\delta_S^* {\cal D}^{ab} \psi^c = \frac{1}{2} \xi^d \psi^e \partial_{de} {\cal D}^{ab}\psi^c,
\eea
and, therefore,
\be\label{trdetE}
\delta_S^* \det{\cal E} = \frac{1}{2} \xi^a {\cal D}_{ab}\psi^b - \frac{1}{8} \xi^d \psi_d {\cal D}^{ab}\psi^c {\cal D}_{ab}\psi_c +
\frac{1}{2} \xi^c \psi^d \partial_{cd} \det{\cal E}.
\ee
Thus, the integrand in the action \p{cact1} transforms as follows:
\bea\label{tract1}
\delta_S^* \Big ( \det{\cal E} {\cal F}\Big )&=& \left( \frac{1}{2} \xi^a {\cal D}_{ab}\psi^b
- \frac{1}{8} \xi^d \psi_d {\cal D}^{ab}\psi^c {\cal D}_{ab}\psi_c\right) {\cal F}
+ \frac{1}{2} \xi^c \psi^d \partial_{cd} \Big ( \det{\cal E}{\cal F}\Big )  \nn \\
&=& \left( \frac{1}{2} \xi^a {\cal D}_{ab} \psi^b - \frac{1}{8} \xi^d \psi_d {\cal D}^{ab}\psi^c {\cal D}_{ab}\psi_c -
\frac{1}{2} \xi^c \partial_{cd}\psi^d \det{\cal E}\right) {\cal F}.
\eea
It is a matter of direct calculations to check that the expression in the parentheses in \p{tract1} is zero. Thus, the action \p{cact1},
as well as the action \p{Action}, are indeed invariant under spontaneously broken supersymmetry.
\subsection*{Unbroken supersymmetry}
It is funny, but in contrast with the superfield approach in which unbroken supersymmetry is manifest, to prove the invariance of the component
action \p{Action} under unbroken supersymmetry is a rather complicated task.

Under unbroken $Q^a$ supersymmetry the covariant derivatives $\nabla_{ab}, \nabla_a$ \p{nabla} are invariant by construction.
Therefore, the objects $\nabla_{ab} \psi_c, \nabla_{ab} q$ are the superfields with the standard transformation
properties
\bea
&& \delta_Q^* \psi^a = -\eta^b (D_b \mpsi^a)|_{\theta=0} = 2 \eta^b\left( \lambda_b{}^a
-\frac{1}{2} \psi^m \lambda_b{}^n\partial_{mn}\psi^a\right), \label{Qpsi} \\
&& \delta_Q^* {\cal D}_{ab} \psi_c = -\eta^d (D_d \nabla_{ab} \mpsi_c)|_{\theta=0}
= -\eta^d\left( 2 {\cal D}_{ab}\psi^m \lambda_d{}^n {\cal D}_{mn}\psi_c
-2{\cal D}_{ab}\lambda_{dc}+
\psi^m\lambda_d{}^n \partial_{mn} {\cal D}_{ab}\psi_c\right), \label{Qnablapsi} \\
&& \delta_Q^* {\cal D}_{ab}q = -\eta^c (D_c \nabla_{ab}\, \boldsymbol{q})|_{\theta=0} =
-\eta^c\left( \frac{1-2\lambda^2}{1+2\lambda^2}{\cal D}_{ab}\psi_c + \psi^m\lambda_c{}^n\partial_{mn} {\cal D}_{ab}\, q \right). \label{Qq}
\eea
Therefore,
\bea\label{QdetE}
\delta_Q^* \det{\cal E} &=& \eta^c\lambda_c{}^a {\cal D}_{ab}\psi^b -\eta^c {\cal D}_{ab}\lambda^b{}_c \psi^a +
\eta^c\lambda_c{}^n \psi^a {\cal D}_{ab}\psi^m {\cal D}_{mn}\psi^b -\frac{1}{4}\, \eta^b\lambda_b{}^a \psi_a {\cal D}^{mn}\psi^k {\cal D}_{mn}\psi_k \nn \\
&-& \frac{1}{8}\, \psi^2 \eta^d \lambda_d{}^b {\cal D}_{bc}\psi^c {\cal D}^{mn}\psi^k {\cal D}_{mn}\psi_k+
\frac{1}{4}\, \psi^2 \eta^d {\cal D}_{ab}\lambda_{dc} {\cal D}^{ab} \psi^c-\eta^c\lambda_c{}^n\psi^m \partial_{mn} \det{\cal E},
\eea
and
\be\label{QF}
\delta_Q^* {\cal F}= -2\, \frac{1-2 \lambda^2}{1+2\lambda^2}\, \eta^c {\cal D}_{ab} \psi_c {\cal D}^{ab}\, q\, {\cal F}'-
\eta^c\lambda_c{}^n \psi^m \partial_{mn} {\cal F}.
\ee
The ${\cal F}'$ in \p{QF} denotes the derivative ${\cal F}$ over its argument (i.e. over ${\cal D} q \cdot {\cal D} q$ in our
case).

Combining these expressions we will get the following variation of the integrand of our action \p{Action}:
\be\label{add1}
\delta_Q^* {\cal L} = \delta_Q^* \Big ( \det{\cal E}\, {\cal F}\Big) = \delta_Q^* \det{\cal E}\, {\cal F} + \det{\cal E}\,\delta_Q^* {\cal F}\,.
\ee
In \p{add1} the last terms from $\delta_Q^* \det{\cal E}$ \p{QdetE} and $\delta_Q^* {\cal F}$ \p{QF} combine together to produce
$$ -\eta^a \lambda_{a}{}^b \psi^c \partial_{bc} \Big ( \det{\cal E}{\cal F}\Big ).$$
Therefore, after integration by parts in this term we will get
\bea\label{add2}
\delta_Q^* {\cal L}&=& \left( \eta^c\lambda_c{}^a {\cal D}_{ab}\psi^b -\eta^c {\cal D}_{ab}\lambda^b{}_c \psi^a +
\eta^c\lambda_c{}^n \psi^a {\cal D}_{ab}\psi^m {\cal D}_{mn}\psi^b-\frac{1}{4}\,\eta^b\lambda_b{}^a \psi_a {\cal D}^{mn}\psi^k {\cal D}_{mn}\psi_k \right.\nn \\
&-&\left. \frac{1}{8}\, \psi^2 \eta^d \lambda_d{}^b {\cal D}_{bc}\psi^c {\cal D}^{mn}\psi^k {\cal D}_{mn}\psi_k+
\frac{1}{4}\, \psi^2 \eta^d {\cal D}_{ab}\lambda_{dc} {\cal D}^{ab} \psi^c\right) {\cal F} \\
&-& 2\, \frac{1-2 \lambda^2}{1+2\lambda^2}\, \eta^c {\cal D}_{ab}\psi_c {\cal D}^{ab} q {\cal F}' \det{\cal E} +
\eta^c \partial_{mn} \lambda_c{}^n \psi^m {\cal F} \det{\cal E} +\eta^c\lambda_c{}^n \partial_{mn} \psi^m  {\cal F} \det{\cal E}\,.\nn
\eea
Now, one may check that terms with the derivatives of $\lambda_{ab}$ in \p{add2} just canceled.

The next step is to substitute into \p{add2} the explicit expressions for $\lambda_{ab}$ \p{ihA} and for ${\cal F}$ \p{Action}
\be\label{add3}
\lambda_{ab}=\frac{-\frac{1}{2} {\cal D}_{ab}\, q}{1+\sqrt{1-\frac{1}{2} {\cal D} q \cdot {\cal D} q}}\,, \qquad
{\cal F}= 1+\sqrt{1-\frac{1}{2} {\cal D} q \cdot {\cal D} q}\,.
\ee
If we note that
\be\label{add4}
\lambda_{ab}=\frac{-\frac{1}{2} {\cal D}_{ab}\, q}{\cal F} \qquad \mbox{and} \qquad
\frac{1-2 \lambda^2}{1+2\lambda^2} =-\frac{1}{4\; {\cal F}'}\,,
\ee
it will be not so strange that after substitution of \p{add3} into \p{add2}, the variation $\delta_Q^* {\cal L}$
will not contain any square roots. So, it will  read
\bea\label{add5}
\delta_Q^* {\cal L}&=& -\frac{1}{2}\eta^c {\cal D}_c{}^a q {\cal D}_{ab}\psi^b -
\frac{1}{2}\, \eta^c {\cal D}_{c}{}^n q \psi^a {\cal D}_{ab}\psi^m {\cal D}_{mn}\psi^b+
\frac{1}{8}\,\eta^b {\cal D}_b{}^a q \psi_a {\cal D}^{cd}\psi^e {\cal D}_{cd}\psi_e \nn \\
&+& \frac{1}{16}\, \psi^2 \eta^a {\cal D}_a{}^b q {\cal D}_{bc}\psi^c {\cal D}^{de}\psi^f {\cal D}_{de}\psi_f+
\frac{1}{2}\, \eta^c {\cal D}_{ab} \psi_c {\cal D}^{ab} q \det{\cal E} \nn \\
&-& \frac{1}{2}\, \eta^a {\cal D}_a{}^b q\; \partial_{bc}\psi^c \det{\cal E}.
\eea
Substituting now the expression for $\partial_{bc}\psi^c \det{\cal E}$ from \p{tract1} and slightly rearranging the terms,
we obtain
\bea\label{add6}
\delta_Q {\cal L}&=& -\eta^c {\cal D}_c{}^a q {\cal D}_{ab}\;\psi^b -
\frac{1}{4}\, \eta^a {\cal D}_{a}{}^b q \;\psi_b {\cal D}^{cd}\psi_d {\cal D}_{ce}\psi^e +
 \frac{1}{16}\, \psi^2 \eta^a {\cal D}_a{}^b q {\cal D}_{bc}\psi^c {\cal D}^{de}\psi^f {\cal D}_{de}\psi_f \nn \\
&+&  \frac{1}{2}\, \eta^c {\cal D}_{ab} \psi_c {\cal D}^{ab} q \det{\cal E}\,.
\eea
Finally, combining the terms in the first line together, we will get the following simple form of the variation of
the integrand
\be\label{varL}
\delta_Q^* {\cal L} = -\eta^c \left( {\cal D}_c{}^a q {\cal D}_{ab} \psi^b-
\frac{1}{2}\, {\cal D}^{ab}\, q {\cal D}_{ab} \psi_c\right) \det{\cal E}\,.
\ee
Unfortunately,  further simplifications are not possible. The simplest way to be sure that $\delta_Q^* {\cal L}$ \p{varL}
gives zero after integration over $d^3x$ is to find the ``equation of motion'' for $q$ which follows from the
``Lagrangian''  \p{varL}
\be\label{add7}
\frac{\delta}{\delta q} \int d^3x \;\delta_Q {\cal L} =0.
\ee
Clearly, the expression \p{add7} has to be identically equal to zero if our action is invariant under unbroken supersymmetry.
After quite lengthly and tedious, but straightforward calculations, one may show that this is indeed so.

Thus, our action \p{Action} is invariant with respect to both broken and unbroken supersymmetries.

\setcounter{equation}{0}
\def\theequation{C.\arabic{equation}}
\section*{Appendix C}
In this Appendix we collected some formulas describing the nonlinear realization
of $N=1, D=5$ Poincar\'{e} group in its coset over $d=3$ Lorentz group $SO(1,2)$.

In $d=3$ notation the $N=1, d=5$  Poincar\'e superalgebra contains the following set of generators:
\be\label{1A}
\mbox{ N=4, d=3 SUSY }\quad \propto \quad \left\{ P_{ab}, Q_a,\bQ_a, S_a,\bS_a, Z, \bZ, M_{ab}, K_{ab}, \bK_{ab}, J \right\},
\ee
$a,b=1,2$ being the $d=3$ $SL(2,R)$
spinor indices \footnote{The indices are raised
and lowered as follows:
$V^{a}=\epsilon^{ab}V_b,\;V_{b}=\epsilon_{bc}V^c,\quad
\epsilon_{ab}\epsilon^{bc}=\delta_a^c\; .$}. Here, $P_{ab}$ $Z$  and $\bZ$ are $D=5$ translation generators,
$Q_a, \bQ_a$ and $S_a, \bS_a$ are the generators of super-translations, the generators $M_{ab}$ form $d=3$
Lorentz algebra $so(1,2)$, the generators $K_{ab}$ and $\bK_{ab}$ belong to the coset $SO(1,4)/SO(1,2)\times U(1)$,
while $J$ span $u(1)$. The basic (anti)commutation relations read
\bea\label{algebra2Z}
&&\left[ M_{ab}, M_{cd}  \right]=\epsilon_{ad}M_{bc}+\epsilon_{ac}M_{bd}+\epsilon_{bc}M_{ad}+\epsilon_{bd}M_{ac} \equiv \left( M\right)_{ab,cd},  \nn \\
&&\left[ M_{ab}, P_{cd}  \right]=\left( P\right)_{ab,cd}, \;
\left[ M_{ab}, K_{cd}  \right]=\left( K\right)_{ab,cd}, \;
\left[ M_{ab}, \bK_{cd}  \right]=\left( \bK\right)_{ab,cd},  \nn \\
&&\left[ K_{ab}, \bK_{cd}  \right] = \frac{1}{2}\left( M\right)_{ab,cd} +2\left( \epsilon_{ac}\epsilon_{bd}
+ \epsilon_{bc}\epsilon_{ad}  \right)J, \nn \\
&&\left[ K_{ab}, P_{cd}   \right] = -\left( \epsilon_{ac}\epsilon_{bd}+  \epsilon_{bc}\epsilon_{ad}  \right) Z, \;
\left[ \bK_{ab}, P_{cd}   \right] = \left( \epsilon_{ac}\epsilon_{bd}+  \epsilon_{bc}\epsilon_{ad}  \right) \bZ,  \nn \\
&& \left[ K_{ab}, \bZ   \right]  = -2P_{ab}, \ \left[ \bK_{ab}, Z   \right]  = 2P_{ab},
\nn\\
&& \left[ M_{ab}, Q_{c}  \right] = \epsilon_{ac}Q_{b}+\epsilon_{bc}Q_{a}, \ \left[ M_{ab}, \bQ_{c}  \right] = \left( \bQ\right)_{ab,c},  \;
 \left[ M_{ab}, S_{c}  \right] = \left( S\right)_{ab,c}, \ \left[ M_{ab}, \bS_{c}  \right] = \left( \bS\right)_{ab,c},  \nn \\
&& \left[ \bK_{ab}, Q_{c}  \right] =- \left( \bS\right)_{ab,c}, \ \left[ K_{ab}, \bQ_{c}  \right] = \left( S\right)_{ab,c},  \;
\left[ \bK_{ab}, S_{c}  \right] = \left( \bQ\right)_{ab,c}, \ \left[ K_{ab}, \bS_{c}  \right] = -\left( Q\right)_{ab,c},  \nn \\
&& \left[ J,Q_{a}\right]=-\frac{1}{2}Q_a,\;\left[ J,\bQ_{a}\right]=\frac{1}{2}\bQ_a,\;
\left[ J,S_{a}\right]=-\frac{1}{2}S_a,\;\left[ J,\bS_{a}\right]=\frac{1}{2}\bS_a,\nn \\
&& \left[ J,K_{ab}\right]=-K_{ab},\;\left[ J,\bK_{ab}\right]=\bK_{ab},\;
\left[ J,Z\right]=-Z,\;\left[ J,\bZ\right]=\bZ,\nn \\
&& \left\{ Q_{a} , \bQ_{b}  \right\} =2P_{ab}, \ \left\{ S_a, \bS_b \right\}=2P_{ab},  \;
\left\{ Q_{a}, S_{b} \right\} =2\epsilon_{ab} Z, \ \left\{ \bQ_{a}, \bS_{b} \right\} =2\epsilon_{ab} \bZ.
\eea
Note, that the generators obey the following conjugation rules:
\bea\label{conrules}
&& \left(P_{ab}\right)^\dagger=P_{ab},\;\left(K_{ab}\right)^\dagger=\bK_{ab},\;\left(M_{ab}\right)^\dagger=-M_{ab},\;
J^\dagger=J,\;Z^\dagger=\bZ,\nn \\
&& \left(Q_{a}\right)^\dagger=\bQ_{a},\;\left(S_{a}\right)^\dagger=\bS_{a}.
\eea
We  define the coset element as follows
\be\label{coset2Z}
g = e^{i x^{ab}P_{ab}}e^{\theta^a Q_a + \bar\theta^a \bQ_a}e^{i(\boldsymbol{q} Z+\bar{\boldsymbol{q}} \bZ)}e^{\mpsi^a S_a + \mbpsi^a \bS_a}
e^{i  (\mLambda^{ab}K_{ab}+ \mbLambda^{ab}\bK_{ab})}.
\ee
Here, $\left\{x^{ab}, \theta^a, \bar\theta^a\right\}$ are $N=2, d=3$ superspace coordinates, while the remaining coset parameters are Goldstone superfields,
$\boldsymbol{q} = \boldsymbol{q}(x,\theta, \bar\theta), \;\bar{\boldsymbol{q}} = \bar{\boldsymbol{q}}(x,\theta, \bar\theta),\;
\mpsi^a = \mpsi^a(x,\theta, \bar\theta),\;\mbpsi^a = \mbpsi^a(x,\theta, \bar\theta),\;
\mLambda^{ab} = \mLambda^{ab}(x,\theta, \bar\theta),\; \mbLambda^{ab} = \mbLambda^{ab}(x,\theta, \bar\theta)$.
These $N=2$ superfields obey the following conjugation rules:
\be\label{conrules1}
\left( x^{ab}\right)^\dagger=x^{ab},\; \left( \theta^a\right)^\dagger =\bar\theta{}^a,\quad
\boldsymbol{q}^\dagger=\bar{\boldsymbol{q}},\; \left( \mpsi^a\right)^\dagger =\mbpsi^a,\; \left(\mLambda^{ab}\right)^\dagger =\mbLambda^{ab}.
\ee
The transformation properties of the coordinates and superfields with respect
to all symmetries can be found by acting from the left on the coset element $g$ \p{coset2Z} by the different elements
of $N=1, D=5$ Poincar\'{e} supergroup. In what follows, we will need only the explicit form only for
the broken $(S, \bS)$, unbroken $(Q,\bQ)$ supersymmetries, and $(K, \bK)$ automorphism transformations which read
\begin{itemize}
\item Unbroken $(Q)$ supersymmetry $(g_0=\mbox{exp }(  \epsilon^{a}Q_{a}+\bar\epsilon^{a}\bQ_{a} ))$
\be\label{susy1A}
\delta x^{ab}=i\, \left( \epsilon^{(a}\bar\theta^{b)}+\bar\epsilon^{(a}\theta^{a)}\right) ,
\quad
\delta \theta^{a}=\epsilon^a\;,\delta \bar\theta^{a}=\bar\epsilon^a\; .
\ee
\item Broken $(S)$ supersymmetry $(g_0=\mbox{exp }\left(  \varepsilon^{a}S_{a}+\bar\varepsilon^{a}\bS_{a} \right))$
\be\label{susy2A}
\delta x^{ab}= i\, \left(\varepsilon^{(a}\mbpsi^{b)}+\bar\varepsilon^{(a}\mpsi^{b)}\right),\quad
\delta \boldsymbol{q}=2i\, \varepsilon_a\theta^a,\quad \delta \bar{\boldsymbol{q}}=2 i\, \bar\varepsilon_a\bar\theta^a,\quad
\delta\mpsi^a=\varepsilon^a,\quad \delta\mbpsi^a=\bar\varepsilon^a\; .
\ee
\item Automorphism $(K,\bK)$ transformations  $(g_0=\mbox{exp } i\,\left(  a^{ab}K_{ab}+{\bar a}{}^{ab}\bK_{ab} \right))$
\bea\label{auto}
&& \delta x^{ab}= -2 i\, \left(a^{ab} \boldsymbol{q} -{\bar a}{}^{ab} \bar{\boldsymbol{q}}\right)-2\theta^c\mpsi_c {\bar a}{}^{ab}+
2\bar\theta{}^c\mbpsi_c a^{ab},\quad \delta\theta^a=-2i\, a^{ab}\mbpsi_b,\quad \delta\bar\theta^a=2 i\, {\bar a}^{ab}\mpsi_b,  \nn \\
&& \delta \boldsymbol{q}=- 2 i\, a^{ab}x_{ab}-2 a^{ab}\left( \theta_a \bar\theta_b - \mpsi_a\mbpsi_b\right),\quad
\delta\mpsi^a=2 i\, a^{ab}\bar\theta_b, \nn \\
&& \delta \bar{\boldsymbol{q}}= 2 i\, \bar{a}^{ab}x_{ab}-2 {\bar a}^{ab}\left( \theta_a \bar\theta_b - \mpsi_a\mbpsi_b\right),\quad
\delta\mbpsi^a=-2 i\, {\bar a}^{ab}\theta_b.
\eea
\end{itemize}

As the next step of the coset formalism, one can construct the Cartan forms
\be\label{CFdef}
g^{-1}d g =\Omega_P+\Omega_Q+\overline\Omega_Q+\Omega_Z+\overline\Omega_Z+\Omega_S+\overline\Omega_S+\ldots\;.
\ee
In what follows we will need only the forms $\left \{\Omega_P, \Omega_{Q}, \Omega_{\bar Q}, \Omega_Z, \Omega_{\bar Z}, \Omega_S, \Omega_{\bar S} \right \}$
which explicitly read
\bea
&&\Omega_P =\left\{ \left(\cosh{2\sqrt{\mY}}\right)_{ab}^{cd}\triangle x^{ab} - i\, \left( \mbLambda^{ab}\triangle \boldsymbol{q}
- \mLambda^{ab}\triangle \bar{\boldsymbol{q}} \right)\left( \frac{\sinh{2\sqrt{\mY}}}{\sqrt{\mY}} \right)_{ab}^{cd} \right\} P_{cd}, \nn \\
&&\Omega_{Q} =\left\{d \theta^{b}  \left(\cos{2\sqrt{\mbT}}\right)_{b}^{\;\, c}- i\, d \mbpsi^{b} \,\mLambda_{b}^{\;\, a}
\left(\frac{\sin{2\sqrt{\mbT}}}{\sqrt{ \mbT}}\right)_{a}^{\;\, c}\right\} Q_c,\nn\\
&& \Omega_Z = \left\{  \triangle \boldsymbol{q}  +  \left( \mbLambda^{ab} \triangle \boldsymbol{q} - \mLambda^{ab}\triangle\bar{\boldsymbol{q}} \right)
\left( \frac{\cosh 2\sqrt{\mY} -1}{\mY} \right) _{ab}^{cd} \mLambda_{cd}
+ i\, dx^{ab} \left( \frac{\sinh{2\sqrt{\mY}}}{\sqrt{\mY}}  \right)_{ab}^{cd} \mLambda_{cd}  \right\}Z, \nn \\
&&\Omega_{S} =\left\{d \mpsi^{b}  \left(\cos{2\sqrt{\mbT}}\right)_{b}^{\;\, c}+ i\, d \bar\theta^{b} \,\mLambda_{b}^{\;\, a}
\left(\frac{\sin{2\sqrt{\mbT}}}{\sqrt{ \mbT}}\right)_{a}^{\;\, c}\right\} S_c,\label{Dx}\\
&&\triangle x^{ab} = d x^{ab} -i \left ( \theta^{(a} d \bar \theta^{b)} + \bar \theta^{(a} d \theta^{b)}
+ \mpsi^{(a} d \mbpsi^{b} + \mbpsi^{(a} d \mpsi^{b} \right )\,,\nn\\
&&\triangle \boldsymbol{q} = d \boldsymbol{q} -2i\, \mpsi_{a}d\theta^{a}, \quad
\triangle \bar{\boldsymbol{q}} = d\bar{\boldsymbol{q}} -2i\, \mbpsi_{a}d\bar\theta^{a} .
\label{Dq}
\eea
Here, we defined matrix-valued functions $\mY_{ab}{}^{cd}, \mT_a{}^b$ and ${\mbT}{}_a{}^b$ as
\be\label{YT}
\mY_{ab}{}^{cd} = \mLambda_{ab}\mbLambda^{cd}+ \mbLambda_{ab}\mLambda^{cd},\quad
\mT_a{}^b=\mLambda_{a}{}^c\mbLambda_{c}{}^b, \;{\mbT}_a{}^b=\mbLambda_{a}{}^c\mLambda_{c}{}^b .
\ee
Note, that all these Cartan forms transform homogeneously under all symmetries.

Having at hands the Cartan forms, one may construct the ``semi-covariant'' (covariant with respect to $d=3$ Lorentz,
unbroken and broken supersymmetries only) as
\be\label{cD1}
{\triangle x}^{ab}\nabla_{ab} +d\theta^a \nabla_a+d\bar\theta^a \bar\nabla_a = dx^{ab} \frac{\partial}{\partial x^{ab}}
+ d\theta^a \frac{\partial}{\partial \theta^a}+d\bar\theta^a \frac{\partial}{\partial \bar\theta^a}.
\ee
Explicitly, they read
\bea\label{nabla1}
&& \nabla_{ab} = (E^{-1})_{ab}{} ^{cd} \partial_{cd},\nn \\
&&\nabla_a = D_a -i\, \left( \mpsi^b D_a \mbpsi^c + \mbpsi^b D_a \mpsi^c  \right)\nabla_{bc} = D_a
- i\, \left( \mpsi^b \nabla_a \mbpsi^c + \mbpsi^b \nabla_a \mpsi^c  \right)\partial_{bc},
\eea
where
\bea
&& D_a=\frac{\partial}{\partial \theta^a} -i\,\bar \theta^b\,\partial_{ab},\;
\bD_a=\frac{\partial}{\partial \bar\theta^a} -i\, \theta^b\,\partial_{ab},\quad
\left\{ D_a, \bD_b\right\} = -2 i\, \partial_{ab}, \label{flatCD} \\
&& E_{ab}{}^{cd} = \delta^{(c}_a \delta^{d)}_b -i\, \left( \mpsi^{(c}\partial_{ab} \mbpsi^{d)} +\mbpsi^{(c}\partial_{ab} \mpsi^{d)}   \right),\label{E2} \\
&& (E^{-1})_{ab}{}^{cd}=\delta^{(c}_a \delta^{d)}_b + i\, \left( \mpsi^{(c}\nabla_{ab} \mbpsi^{d)} +\mbpsi^{(c}\nabla_{ab} \mpsi^{d)}   \right) \label{Em1}.
\eea
The derivatives obey the following algebra:
\bea\label{deralg}
&&\left\{\nabla_a, \nabla_b\right\}= -2i\, \left( \nabla_a \mpsi^c  \nabla_b \mbpsi^d + \nabla_a \mbpsi^c  \nabla_b \mpsi^d  \right)\nabla_{cd} , \nn \\
&&\left\{\nabla_a, \bnabla_b\right\}=-2i\, \nabla_{ab} -2i\, \left( \nabla_a \mpsi^c  \bnabla_b \mbpsi^d
+ \nabla_a \mbpsi^c  \bnabla_b \mpsi^d  \right)\nabla_{cd} , \nn \\
&&\left[\nabla_{ab},\nabla_c\right] = -2i\, \left( \nabla_{ab} \mpsi^d  \nabla_c \mbpsi^f + \nabla_{ab} \mbpsi^d  \nabla_c \mpsi^f  \right)\nabla_{df} ,\nn\\
&& \left[\nabla_{ab}, \nabla_{cd}\right] =2i\, \left( \nabla_{ab} \mpsi^m \nabla_{cd} \mbpsi^n
- \nabla_{cd} \mpsi^m \nabla_{ab} \mbpsi^n \right) \nabla_{mn}.
\eea
 The $d=3$ volume form is defined as
\be\label{volume}
d^3 x \equiv \epsilon_{IJK} dx^I \wedge dx^J \wedge dx^K \quad \Rightarrow \quad
dx^I \wedge dx^J \wedge dx^K = \frac{1}{6} \epsilon^{IJK} d^3 x.
\ee
Transition from the spinor notations to the vector one is set as follows
\be\label{44}
V^I \equiv \frac{i}{\sqrt{2}}\left(\sigma^I\right)_a{}^b\; V_b{}^a \quad \Rightarrow \quad V_a{}^b
= -\frac{i}{\sqrt{2}} V^I \left( \sigma^I\right)_a{}^b,  \qquad V^{ab}V_{ab} = V^I V^I.
\ee
Here we are using the standard set of $\sigma^I$ matrices
\be\label{41}
\sigma^I \; \sigma^J = i\, \epsilon^{IJK} \sigma^K + \delta^{IJ} {\cal I}\,, \quad
\left( \sigma^I\right)_a{}^b \; \left(\sigma^I \right)_c{}^d = 2 \delta_a{}^d \delta_c{}^b - \delta_a{}^b \delta_c{}^d,
\ee
were $\epsilon^{IJK}$ obeys relations
\be\label{42}
\epsilon^{IJK}\epsilon_{IMN} =\delta^{J}_{M}\delta^{K}_{N} -\delta^{J}_{N}\delta^{K}_{M},\quad
\epsilon^{IJK}\epsilon_{IJN}=2 \delta^{K}_{N}, \quad \epsilon^{IJK}\epsilon_{IJK}=6.
\ee

\end{document}